\documentclass[11pt,letterpaper]{article}
\usepackage[T1]{fontenc}
\usepackage[utf8]{inputenc}
\usepackage{lmodern}
\usepackage{upgreek}
\usepackage{textcomp}
\usepackage{amsmath,amsfonts,amssymb,latexsym,amsthm,mathrsfs}
\usepackage[english]{babel} 
\usepackage{indentfirst,setspace,fancyhdr}
\usepackage[outercaption]{sidecap}
\usepackage[footnotesize]{caption}
\usepackage[margin=3cm]{geometry}
\usepackage[pdftex]{graphicx}
\usepackage{color,xcolor}
\usepackage[none]{hyphenat}
\usepackage[toc,page]{appendix}
\usepackage{colortbl}
\usepackage{gensymb}
\usepackage{multirow}
\usepackage{array}
\usepackage{pstricks}
\usepackage{wrapfig}
\usepackage{pdfpages}
\usepackage{mathtools}
\usepackage{booktabs} % Better table styles
\usepackage{cite}
\usepackage{quoting}
\usepackage{longtable}
\usepackage{cancel}
\usepackage{makecell}
\usepackage{tikz,stackengine}
\usetikzlibrary{matrix,calc}
\usepackage{blkarray}
\usetikzlibrary{decorations.pathreplacing}

\usepackage{tabularx}
\usepackage{authblk}

\usepackage{setspace}

\usepackage{soul}
\usepackage[normalem]{ulem}

\usepackage{abstract}
\addto{\captionsenglish}{}

\newcommand{\xddots}{%
	\raise 4pt \hbox {.}
	\mkern 6mu
	\raise 1pt \hbox {.}
	\mkern 6mu
	\raise -2pt \hbox {.}
}
\emergencystretch=\maxdimen
\DeclareFontFamily{OT1}{pzc}{}
\DeclareFontShape{OT1}{pzc}{m}{it}{<-> s * [1.10] pzcmi7t}{}
\DeclareMathAlphabet{\mathpzc}{OT1}{pzc}{m}{it}

\graphicspath{{Figures/}} 
\newcolumntype{g}{>{\columncolor{light-gray}}c}

\usepackage{sectsty}
\allsectionsfont{\sffamily}

\usepackage[labelformat=empty, position=top]{subcaption}
\definecolor{Gray}{rgb}{0.5,0.5,0.5}
\definecolor{darkblue}{rgb}{0.12,0.15,0.49}
\definecolor{light-gray}{gray}{0.85}
\definecolor{mygreen}{rgb}{0, 0.5, 0}
\makeatletter
\renewcommand\@biblabel[1]{#1.} 
\makeatother

\usepackage{xr}
\usepackage{setspace}
\usepackage{filecontents}

\title{\bf \sffamily Mito-nuclear selection induces a trade-off between\\ species ecological dominance and evolutionary lifespan}
\author[1]{Débora Princepe}
\author[1]{Marcus A. M. de Aguiar}
\author[2]{Joshua B. Plotkin}
\affil[1]{Instituto de F\'{i}sica `Gleb Wataghin', Universidade Estadual de Campinas, Campinas, Brazil}
\affil[2]{Department of Biology, University of Pennsylvania, Philadelphia, PA, USA}
\date{}

\begin{document}

\doublespacing 
\selectlanguage{english}
\setlength{\parindent}{0cm}
\maketitle

%\begin{center}
%\textbf{\Large \sffamily Mito-nuclear selection induces a trade-off between\\ species ecological dominance and evolutionary lifespan}
%
%\vspace{.2cm}
%
%Débora Princepe$^1$, Marcus A. M. de Aguiar$^1$ and Joshua B. Plotkin$^2$\end{center}
%
%\centerline{ $^1$ Instituto de F\'isica, Universidade Estadual de Campinas, Campinas, SP, Brazil}
%
%\centerline{$^2$Department of Biology, University of Pennsylvania, Philadelphia, PA, USA}
%
%\vspace{0.5cm}
\begin{abstract}

\noindent 
Mitochondrial and nuclear genomes must be co-adapted to ensure proper cellular respiration and energy production. Mito-nuclear incompatibility reduces individual fitness and induces hybrid infertility, suggesting a possible role in reproductive barriers and speciation. 
%Then, for being linked to metabolic and ecological characteristics, such a mechanism leaves signatures within individuals, intraspecies and interspecies. 
Here we develop a birth-death model for evolution in spatially extended populations under selection for mito-nuclear co-adaptation. Mating is constrained by physical and genetic proximity, and offspring inherit nuclear genomes from both parents, with recombination. The model predicts macroscopic patterns including a community's long-term species diversity, its species abundance distribution, speciation and extinction rates, as well as intra- and inter-specific genetic variation. We explore how these long-term outcomes depend upon the  microscopic parameters of reproduction: individual fitness governed by mito-nuclear compatibility, constraints on mating compatibility, and ecological carrying capacity. 
%
%
%We explore the connection between  micro-evolutionary processes and ecological and evolutionary patterns, unveiling how a species' longevity is associated with its commonness or rarity and the impact of selection on patterns of speciation and extinction. 
We find that strong selection for mito-nuclear compatibility reduces the equilibrium number of species after a radiation, increases the species' abundances, while simultaneously increasing both speciation and extinction rates. 
%
%total number of species Intraspecific diversity increases as species become more abundant, leading to frequent evolutionary branching that reduces species lifetimes. In equilibrium, such communities present high rates of speciation, extinction, and hybridization, i.e., high turnover rate. 
The negative correlation between species diversity and diversification rates in our model agrees with the broad empirical pattern of lower species diversity and higher speciation/extinction rates in temperate regions,  compared to the tropics. We therefore suggest that these empirical patterns may be caused in part by latitudinal variation in metabolic demands, and corresponding variation in selection on mito-nuclear function.

\end{abstract}

\newpage
\setlength\parindent{1cm} 
\section{Introduction}

The origin, maintenance, and extinction of species involve diverse ecological and evolutionary mechanisms, so that they are rarely explained by a single hypothesis \cite{Hagen2021, Urban2020}. Mathematical models have still been helpful in understanding empirical patterns of biodiversity, such as species abundance distributions \cite{Hubbell2001, Volkov2003, McGill2007} and the species-area curve \cite{aguiar_global_2009, ODwyer2010}, which are well described by both neutral and niche-based theories of community assembly \cite{Chisholm2010, Mutshinda2009}. However, models that explain these ecological features in a community seldom make predictions for long-term evolutionary patterns, such as speciation rates \cite{Rosindell2010} and species lifetimes \cite{Chisholm2014, Nee2005, Ricklefs2003}, nor do they provide a mechanistic description for the formation of species \cite{Etienne2007b, Davies2011}. Studies describing mechanisms of speciation, on the other hand, using both genetic \cite{higgs_stochastic_1991,Gavrilets1998} and ecological \cite{dieckmann_origin_1999,Gavrilets2005,Nosil2012} drivers, rarely connect these processes to macro-evolutionary patterns \cite{Gavrilets2000,Costa2018,Li2018,Melian2012}, whereas macro-ecological models succeed in reproducing such patterns \cite{Rangel2018,Pontarp2019} but may lack a detailed connection with organism and population properties to explain variation in diversification rates \cite{Alencar2021,Hurlbert2014}. %In this context, understanding the connection between ecological and evolutionary aspects in assembling biodiversity can help to explain phenomena that depend on the interaction between them \cite{Ezard2011}, for example, how species duration is associated with its commonness or rarity \cite{Rosindell2015, Thuiller2013}. 
In this context, linking micro-evolutionary processes to macro-evolutionary patterns is key to understanding what drives diversification 
%and identify signatures that these processes could imprint at the 
\cite{Morlon2014,Rosindell2015,Rosindell2013,Etienne2012,Rabosky2013}, especially when similar evolutionary signatures can result from different population processes \cite{Li2018, Costa2018,Louca2020}. 

% Linking micro-evolutionary processes to macro-evolutionary patterns is key to understand the mechanisms driving diversification and identify signatures that these processes could imprint at the evolutionary scale \cite{Morlon2014,Rosindell2015}. This is especially true in light of evidence that similar evolutionary signatures can result from different processes in populations \cite{Li2018, Costa2018}, and that generalized birth-death models can produce extant trees that correspond to countless different diversification processes \cite{Louca2020}. On this basis, models that explicitly include a micro-macro link may provide explanations for ecological patterns \cite{Rosindell2013} while also elucidating signatures in phylogenetic trees \cite{Etienne2012,Costa2018}, for instance. Furthermore, population genetics can help elucidate how these processes connect across different scales \cite{Vellend2005}, such as how selection at the individual level can lead to genetic diversity and evolution of reproductive barriers \cite{Cutter2016, Rabosky2013, Harvey2017,Smith2017}.

Here we study spatially distributed populations that evolve under selection for mating compatibility between parents as well as mito-nuclear compatibility within individuals \cite{Princepe2021}. Genetic interactions between the mitochondrial (mtDNA) and nuclear (nDNA) genomes is necessary for cellular respiration, which depends on biochemically compatible proteins \cite{Bar-Yaacov2012, Sunnucks2017}. Incompatibilities between those genes can severely reduce individual fitness and fertility \cite{Hill2017, Lima2019}. Thus, mutations in the mitochondrial genome induce selection for compatible genes in the nuclear genome, driving molecular co-evolution \cite{Barreto2013} and ultimately promoting compensatory changes \cite{Hill2020}. Mito-nuclear co-evolution has been hypothesized as a driver of speciation \cite{Gershoni2009, Hill2016, Tobler2019} possibly preceding the evolution of reproductive barriers \cite{Burton2012, Barreto2013, Telschow2019}, although its extent in nature remains unknown \cite{Lane2009}. The influence of the respiration efficiency in the availability of energy in eukaryotes, which connects to ecological traits \cite{Hill2019a} and metabolic constraints \cite{Dowling2007}, suggests that mito-nuclear interaction leaves signatures across biological scales, from cellular metabolism to individuals and populations \cite{Sunnucks2017,Wolff2014,Koch2021}. %Therefore, the mito-nuclear interaction provides a natural connection between population genetics, ecology, and evolutionary processes.% that must be further explored. 

We develop an individual-based model to describe selection on mito-nuclear compatibility and we follow the process of radiation and eventual equilibrium towards a stationary diversity of species. We incorporate the mitochondrial genetic material and its interaction with a portion of the nuclear genome in a phenomenological approach \cite{Princepe2021} based on the lock-and-key principle that guides the interaction between proteins coded by both genomes \cite{Wolff2014, Hill2017}. Simulations of our model record all forms of diversification events -- speciation, extinction, and hybridization -- and evaluate the impact of selection on the emergence and persistence of species. In addition, we can use the model to reconstruct the history of an extant community, analyzing the dominance of species throughout their evolutionary lifetime. The model reveals how mito-nuclear selection can jointly influence species abundances and diversity in a community as well as long-term patterns of species lifetimes.

We use the results of our model to help explain longitudinal patterns in species diversity and speciation rates, related to a cline of environmental harshness. Selection on mito-nuclear compatibility is expected to be stronger in harsh, temperate environments that demand optimal metabolic function. Whereas tropical environments, where resources are abundant and energy efficiency is less important, present weaker selection on mito-nuclear compatibility and function. There are well-known patterns in diversity from tropical to temperate regions, with the former displaying high biodiversity, old species, and low speciation and extinction rates, and the latter with fewer species and elevated recent speciation rates \cite{ Weir2007, Botero2014, Weir2014, Schluter2017, Harvey2020, Rabosky2018}. We hypothesize that mito-nuclear selection might play a role in forming these patterns, as resources availability should correlate to selection on metabolic efficiency which, in turn, governs patterns of diversity and speciation rates in our model.

\section{Results}

\subsection{Model for mito-nuclear interaction}

\begin{figure}[!ht]
\begin{center}
    \includegraphics[width=9.5cm]{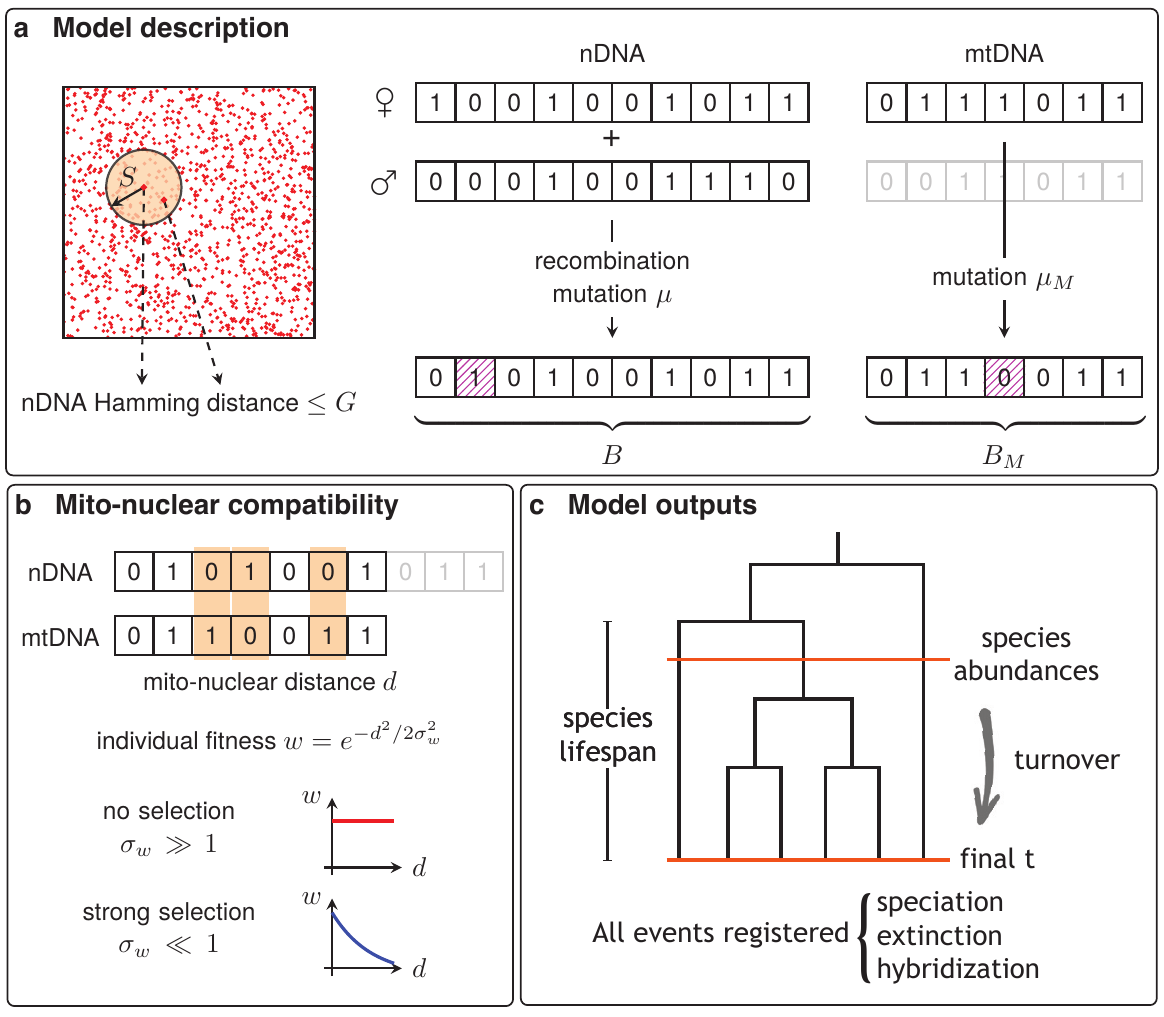}
\end{center}
\caption{\textbf{Model of selection on mito-nuclear compatibility.} \textbf{a}, Individuals are randomly distributed in a square domain and carry nuclear (nDNA) and mitochondrial (mtDNA) genomes  described by binary sequences of $B$  and $B_M$ sites, respectively. Sexual reproduction occurs with non-overlapping generations. Individuals look for a mating partner within a spatial range of radius $S$, and they must be genetically compatible (within nuclear distance $G$) to mate. Offspring are generated with recombination of the nDNA and mutation probability $\mu$ per locus, with direct inheritance of maternal mtDNA with mutation $\mu_M$. \textbf{b}, Mito-nuclear compatibility of an individual is measured by the distance between its mtDNA and $B_M$ corresponding within the nDNA. In the absence of mito-nuclear selection, all individuals have equal fitness. Under selection for mito-nuclear compatibility fitness is assigned to an individual according to its mito-nuclear distance, with strength of selection $\sigma_w$, with higher fitness implying higher probability to reproduce. \textbf{c}, Species are classified as groups of individuals with ongoing gene flow, and their abundances are recorded at each generation. We register all speciation, hybridization, and extinctions events, and record the lifespans of all species.} 
\label{fig:model}
\end{figure}

We employed an agent-based model (Fig. \ref{fig:model}, Methods) based on \cite{aguiar_global_2009}, in which neutral processes driven by recombination and mutation change the allele frequencies in a population. The environment is homogeneous and the population size is kept constant at the carrying capacity. Mating is restricted by both genetic similarity and spatial proximity, so that species emerge through isolation by distance \cite{Costa2018}. Here we start from the same neutral assumptions and introduce the mito-nuclear genetic compatibility as a measure of fitness of individuals. Simulating populations that evolve under these micro-evolutionary dynamics, we catalogue the resulting macro-evolutionary patterns under weak and strong mito-nuclear selection. 

Following \cite{Princepe2021}, each individual is described by a haploid, bi-allelic sequence of $B$ sites that represent its nuclear genome (nDNA). To describe the mito-nuclear interaction we also consider a shorter string of $B_M$ sites representing the individual's mitochondrial genome (mtDNA). Sexes are separated, and sexual reproduction with non-overlapping generations occurs as follows: during the reproduction phase, all individuals have a probability to randomly choose a mating partner within a spatial range of radius $S$, called the {\it mating neighborhood.} Genetic compatibility between individuals is determined by the Hamming distance between their nDNA's, i.e., the number of loci bearing different alleles, and must be below the threshold of $G$ sites for successful mating. The nuclear genome of the offspring is built from recombination of the nuclear material of the parents, with a probability of mutation $\mu$ per site. The mtDNA is directly inherited from the mother, with mutation probability per site $\mu_M$. Starting from identical individuals, species emerge as clusters of genetically compatible individuals, i.e., as groups with ongoing gene flow among individuals, determined by the genetic mating restriction $G$, but no gene flow between different groups (see Methods Section 4.1).

We model mito-nuclear compatibility in an individual as a locus-by-locus interaction between the mtDNA and the first $B_M$ sites of the nuclear genome (the remaining $(B-B_M)$ nuclear genes do not participate in the interaction); the interacting pair of loci is compatible if they have the same allele value (Fig. \ref{fig:model}b). The mito-nuclear distance $d$ is then the fraction of incompatible sites. Such a scheme is a phenomenological description of the biochemical and structural compatibility between proteins coded by both genomes necessary for the respiration process \cite{Sunnucks2017,Wolff2014}. In the absence of selection, all individuals have an equal probability of reproduction, and nDNA and mtDNA evolve independently. When mito-nuclear compatibility is under selection, fitness $w$ is assigned to individuals accordingly to their mito-nuclear distance, with strength expressed by a parameter $\sigma_w$, such that $w=e^{-d^2/2\sigma_w^2 }$, i.e., small $\sigma_w$ implies strong selection. Individuals with higher fitness, normalized within the mating range, have a higher probability of reproduction. We tracked all events during the dynamics, registering speciation, hybridization, and extinctions, and recorded the ancestry of all species. 

\subsection{Diversification rates and species richness}

\begin{figure}[ht]
\begin{center}
    \includegraphics[width=9.5cm]{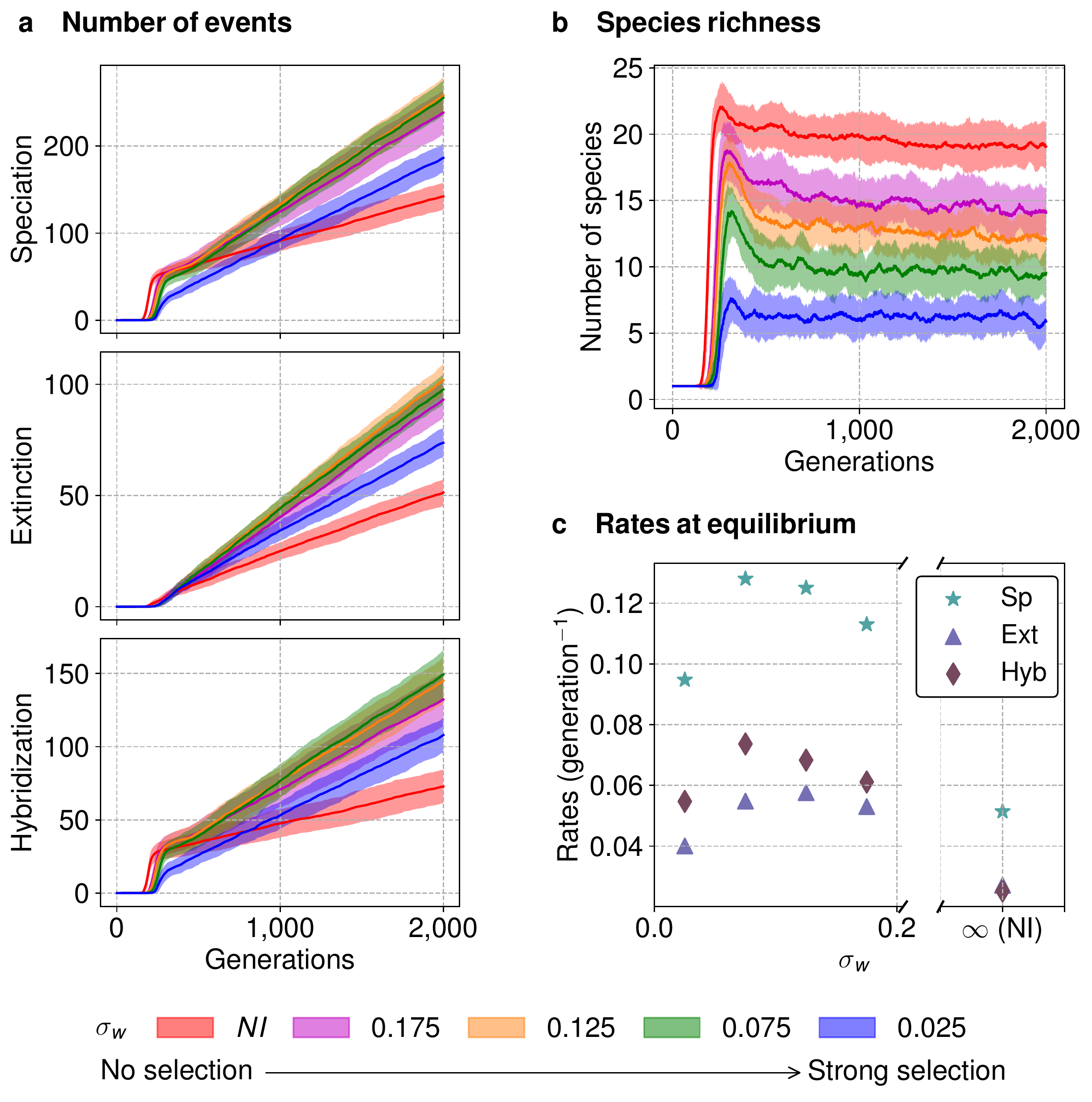}
\end{center}
	\caption{\textbf{Diversification events and species richness.} \textbf{a}, Accumulated number of speciation, extinction, and hybridization events over time. \textbf{b}, Species richness as function of time. \textbf{c}, Rates of speciation (Sp), extinction (Ext), and hybridization (Hyb) obtained from regression of \textbf{a} in equilibrium ($T>500$). Results are shown for varying strength of the mito-nuclear selection $\sigma_w$ (smaller $\sigma_w$  implies stronger selection) and in the absence of selection (NI). Solid lines represent the ensemble mean over 50 independent realizations; and the shadowed area represents one standard deviation from the mean. Selection decreases the number of species at equilibrium, but it increases all rates of diversification events in steady-state, with maximum rates occurring at intermediary values of $\sigma_w$.}
	\label{fig:acc}
\end{figure}

We simulated the evolution of populations under various strengths of selection, $\sigma_w$. In the non-interacting scenario (NI), the mito-nuclear distance is not considered for reproduction (equivalent to $\sigma_w \rightarrow \infty$). Selection was simulated with $\sigma_w$ ranging from 0.175 (weak selection) to 0.025 (strong selection). Results are shown for 50 independent realizations for each set of parameters. Starting from a population of identical individuals, all simulations are characterized by a period of rapid increase in species numbers (a radiation period), followed by relaxation to a steady-state species richness. During the dynamics, all diversification events were registered (Fig. \ref{fig:acc}a). The populations were simulated for $T=2,000$ generations, a time longer than necessary for the completion of the radiation. The equilibrium, defined by the moment when the number of species reaches a plateau, occurred after about $T=500$ generations in all cases (Fig. \ref{fig:acc}b). Because the population size is nearly constant, equilibrium results from a balance between speciation, extinction and hybridization events (compatible with the hypothesis of ecological limits \cite{Rabosky2015}). Selection had the clear effect of reducing the number of species in equilibrium and introducing a delay in the radiation process. 

We define a speciation event, or branching, as the moment when a species splits into two or more reproductively isolated groups. The lifetime of the mother species is considered over at this point and the species disappears as two (or more) sister species are born, with a positive net balance in the number of species. Species also disappear by extinction, when the population becomes very small and eventually does not produce descendants due to random ecological drift and by the accumulation of mito-nuclear incompatibilities that lead to low fitness and low fecundity; or by hybridization when two or more species reestablish gene flow. When species hybridize, we consider that the more abundant species persists while the other one disappears as it merges. Linear regression of the number of events in equilibrium provided the rates of speciation, extinction, and hybridization (Fig. \ref{fig:acc}c). In the equilibrium, when the average number of species becomes constant, the rate at which species are born (speciation) is equal to the rate at which species die (extinction plus hybridization). In this regime, speciation, extinction and hybridization rates per species increase with the strength of selection (Fig.  \ref{fig:ratio}). However, since the number of extant species decreases with selection, the nominal rates have a maximum for intermediate values of $\sigma_w$, as shown in Figure \ref{fig:acc}c.
The decrease in species richness, on the other hand, results from the radiation process, because selection decreases the net diversification (speciation minus extinction and hybridization) during the transient (Fig. \ref{fig:div_rates}).

\subsection{Species abundance distributions and lifespans} 

\begin{figure}[!b]
\begin{center}
        \includegraphics[width=0.8\linewidth]{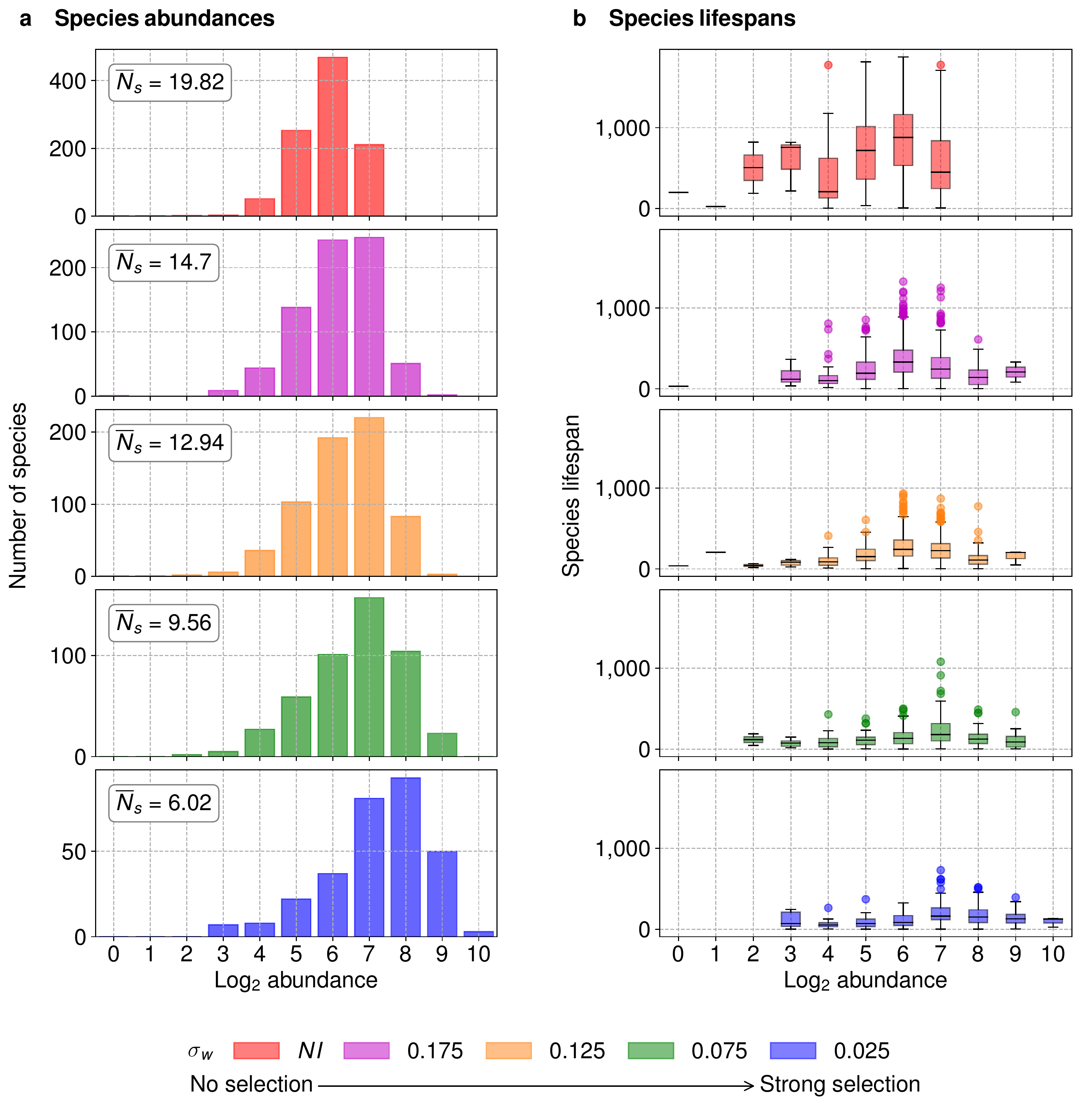}
\end{center}
	\caption{\textbf{Species abundances distributions (SADs) at $T=1,000$ and respective lifespans. }  \textbf{a}, SADs obtained from simulations. Species are binned into log2 abundance categories, following Preston \cite{Gray2006}. Data were accumulated from 50 independent simulations with population size $M=1,300$. $\bar{N}_s$ is the mean number of species. \textbf{b}, Respective species' lifespans within each abundance bin. Boxplots show the median, lower, and upper quartiles, with whiskers extending to 1.5 times the interquartile range; all other points are plotted as outliers. Selection increases mean species' abundances and the variance of the SADs, but it tends to reduce all species lifetimes. At any point of observation, species of intermediate abundance have the longest mean lifetimes. }
	\label{fig:sad}
\end{figure}

Selection on mito-nuclear compatibility decreased the number of species in equilibrium, and because the population size is fixed, this led to an increase in the average species abundance and variance of the species abundance distributions (SADs) (Fig. \ref{fig:sad}a). Since our simulations resulted in an average number of species no greater than 25, we accumulated data over several independent runs to obtain a clear signal for the expected distribution of abundances, for each value of $\sigma_w$. This procedure is justified because the lattice is much larger than the mating range and species' range size are smaller than the available space (see Supplementary Information, Section \ref{SM2}). 

We recorded species abundances in generation $T=1,000$ and found a roughly lognormal shaped distribution of species abundances, even in the presence of strong selection on mito-nuclear compatibility (Fig. \ref{fig:sad}a). Selection increased both the mean species abundance and the variance: the histograms became wider and right-shifted under strong selection (Fig. \ref{fig:sad-skew}). The SADs are asymmetric, and, although they present a tail on the left side, low-abundant species were infrequent because speciation occurs in our model by fission of abundant species, producing relatively few rare species (similarly to a stick breaking speciation model in a community with finite resources \cite{Sugihara1980}). SADs were stable during equilibrium, such that distributions taken at any time step after $T>500$ generations were very similar (Fig. \ref{fig:sad-time}).

The time of birth and disappearance of all species were also recorded, then species age and remaining lifespan were known for all species present at time $T=1,000$ (Fig. \ref{fig:remain}).
The distribution of species lifespans with species binned by same abundance class (Fig. \ref{fig:sad}b) shows that selection tended to reduce species ages and lifetimes in all abundances classes. In all cases, species born with intermediary relative abundance had the highest longevity within their communities. In the absence of selection (NI), species lifetimes were short for low-abundant species and were equally distributed between species of intermediate or large abundance. 
Some of those species persisted until the end of the simulation ($T=2,000$). Surprisingly, large species evolving under strong selection showed drastically reduced longevity. 
In the non-interacting case (no selection), the most-right filled bin (log2 abundance class 7) had a slight reduction of lifespan, indicating that there was also a small effect due to species abundance, although selection seems to be the greater cause of lifespan reduction. 
With regards to species ages at the time of sampling ($T=1,000$) and their remaining lifetimes, the patterns were similar: species of intermediate abundance are the oldest and have the longest remaining lifetimes and species' ages of any abundance class decrease with selection (Fig. \ref{fig:remain}). 

\begin{figure}[h]
\begin{center}
     \includegraphics[width=0.42\linewidth]{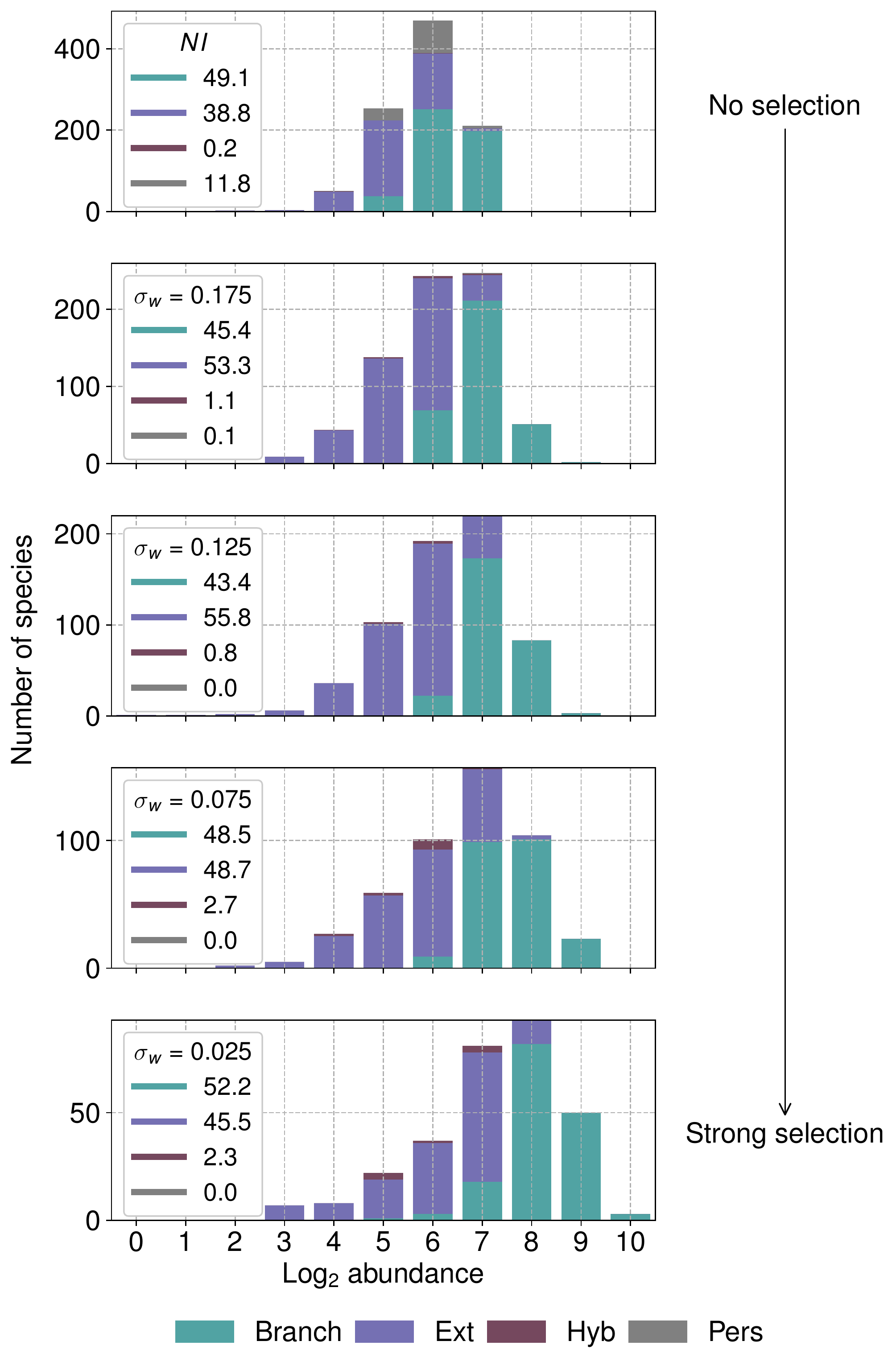}   
\end{center}
	\caption{\textbf{Long-term evolutionary fates of species that were extant at time $T=1,000$.} The histograms depicts the fate of each species shown in Fig. \ref{fig:sad}a at time: by branching (Branch), extinction (Ext), or hybridization (Hyb), or persistence through time $T=2000$ (Pers), binned by log2 abundance class. The legends indicate the percentages of total species that correspond to each event category. Most species die by branching if they are dominant within the community, or by extinction if they are rare. Selection for mito-nuclear compatibility increases rates of hybridization and suppresses long-term persistence.  }
	\label{fig:fate}
\end{figure}

Species abundances at time $T=1,000$ were also related to the distribution of subsequent diversification events (Fig. \ref{fig:fate}), i.e., the mode of disappearance of each species (branching, extinction, or hybridization) or whether it persisted through generation $T=2,000$. Regardless of selection strength, most species had two possible endings: by branching, if it was dominant within the community, or by extinction, if it had low abundance. Remarkably, in the absence of mito-nuclear selection, a significant portion of species persisted through to generation $T=2,000$. Selection suppressed the persistence of species, even when weak ($\sigma_w=0.175$). Selection also increased the rate of hybridization (note that hybridization events were counted only for low-abundant species due to our convention of considering the least abundant species involved as dying during the fusion, whereas the more abundant species persists with increased abundance). 
Notably, species belonging to the same class of abundances had different fates, depending on whether they were under strong, weak, or no selection. For instance, a species belonging to bin 6 in the non-interacting scenario could persist for 1,000 generations ahead, while a species of the same abundance was most likely to go extinct in any scenario with selection. 
Whereas abundance class 7 was the largest one in simulations without selection, more likely to speciate, but it was not dominant and more likely to go extinct under strong selection.

Greater species abundances are expected when the number of species is smaller, because we assume a constant population size; but it is not clear if the macro-evolutionary patterns we observe are attributable to species abundances per se, or to mito-nuclear selection. To tease apart these mechanisms, we simulated populations in the absence of mito-nuclear selection, relaxing the genetic threshold for mating compatibility (from $G=75$ to $G=220$) so that they produced a similar number of species in steady-state as communities under strong mito-nuclear selection (Fig. \ref{fig:g220-abund}). Species-abundance distributions were shifted towards higher abundances as expected, but all diversification rates were reduced, and species lifespans were significantly longer (hence turnover was not increased), even though radiation was delayed (Fig. \ref{fig:g220-nsp}). %Species fates were also very distinct: they are much more likely to persist in the absence of selection (Fig. \ref{fig:g220-events}). 
Therefore, we conclude that the relationship between high speciation rate and low species diversity results directly from the selection on mito-nuclear compatibility.

\subsection{Genetic diversity of the populations}

\begin{figure}[!b]
\begin{center}
    \includegraphics[width=0.9\linewidth]{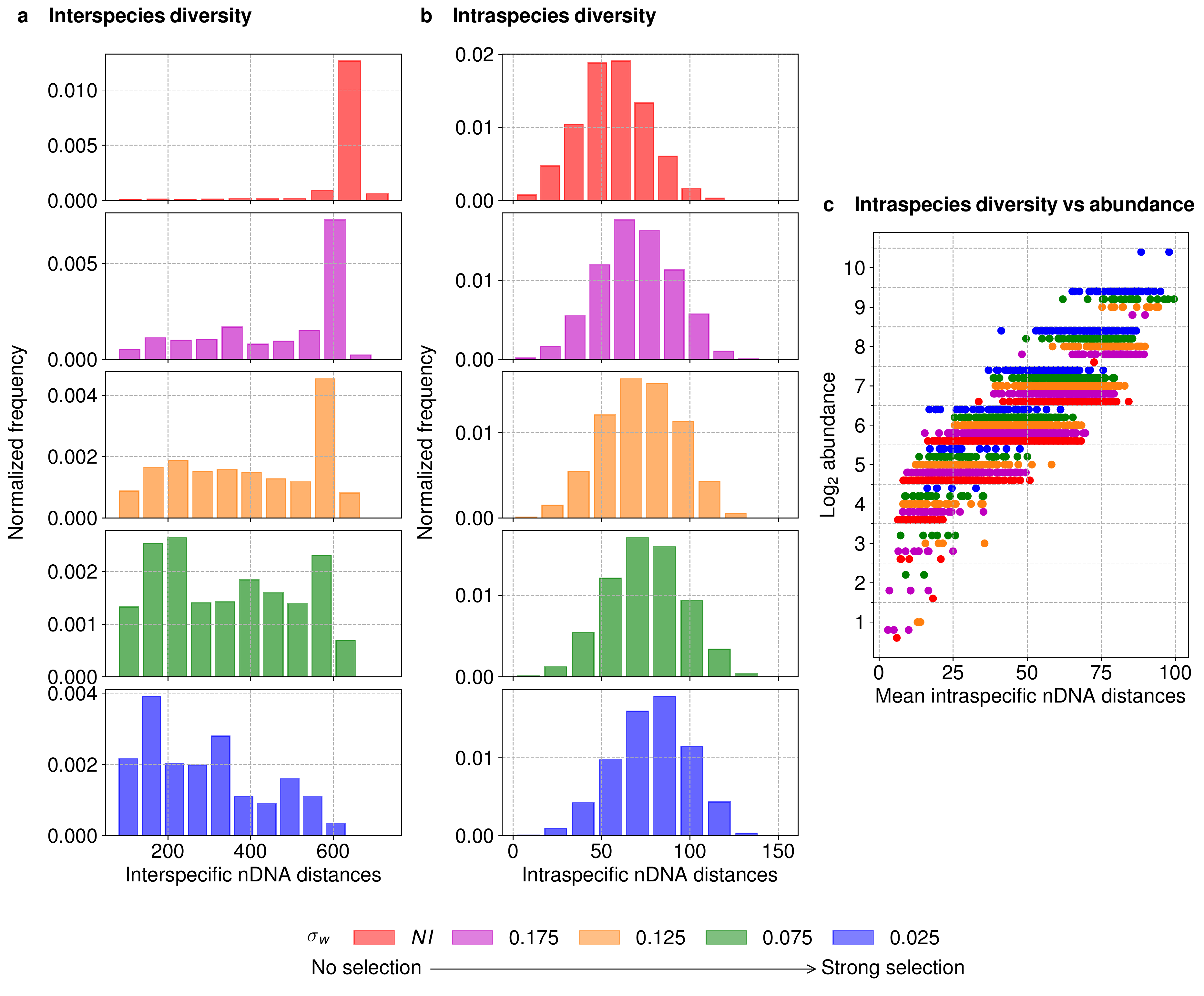}
\end{center}
	\caption{\textbf{Genetic diversity within and between species.} The histograms depict the nDNA distances between pairs of individuals that belong to different species (\textbf{a}), or are conspecific (\textbf{b}) after $T=2,000$ generations. We also show the mean nDNA distance within each species, plotted against the species abundance classes (\textbf{c}). nDNA distances between different species are much lower under strong selection, while the nDNA distances within species are slightly higher. The latter effect is largely due to species abundances: as shown in \textbf{c}, more abundant species allow for greater intraspecific diversity, independent of selection strength.  }
\label{fig:dist}
\end{figure}

The analysis of genetic patterns can help to elucidate the mechanism by which mito-nuclear selection promotes or hinders diversification. We evaluated genetic diversity by measuring the nuclear genetic distance between pairs of individuals at time $T=2,000$. Mito-nuclear selection drastically reduced the genetic distances between individuals belonging to different species (Fig. \ref{fig:dist}a). In the absence of selection (NI), species were genetically well isolated from each other, with a prominent peak around 650 loci. Increasing the strength of selection (reducing $\sigma_w$) made species progressively more similar to each other genetically. However, the effect was opposite for conspecific individuals (Fig. \ref{fig:dist}b): under selective pressure, we observed a slightly larger nuclear divergence within a species than in the non-selective case.  Notably for strong selection, several pairs of individuals had distances exceeding the genetic threshold ($G=75$) required for mating. In this case, species cohesion was kept by ongoing gene flow.  

The patterns of diversity between different species were similar for the mtDNA content (Fig. \ref{fig:mtdna}a) and is explained by the effect of the mito-nuclear selection on reducing both nuclear and mitochondrial substitutions. When reproduction can only occur in individuals with high fitness (low mito-nuclear distance), fixation of a mutation in any of the genome is hampered, because it must be followed by a mutation in the corresponding site of the interacting pair to maintain the mito-nuclear compatibility.  As a result, interacting sites had lower substitution rates in both the mitochondrial and nuclear genomes (Fig. \ref{fig:subs}).  Within a species, the mean nDNA distance is positively correlated to the species' abundance (Fig. \ref{fig:dist}c), independent of the selection strength.  As selection increased species abundances, the effect on the population was to increase the nDNA distances within each species.  Therefore, the mito-nuclear interaction promoted the intraspecies genetic diversity but reduced the global diversity in the community.

\begin{figure}[!h]
\begin{center}
    \includegraphics[width=\linewidth]{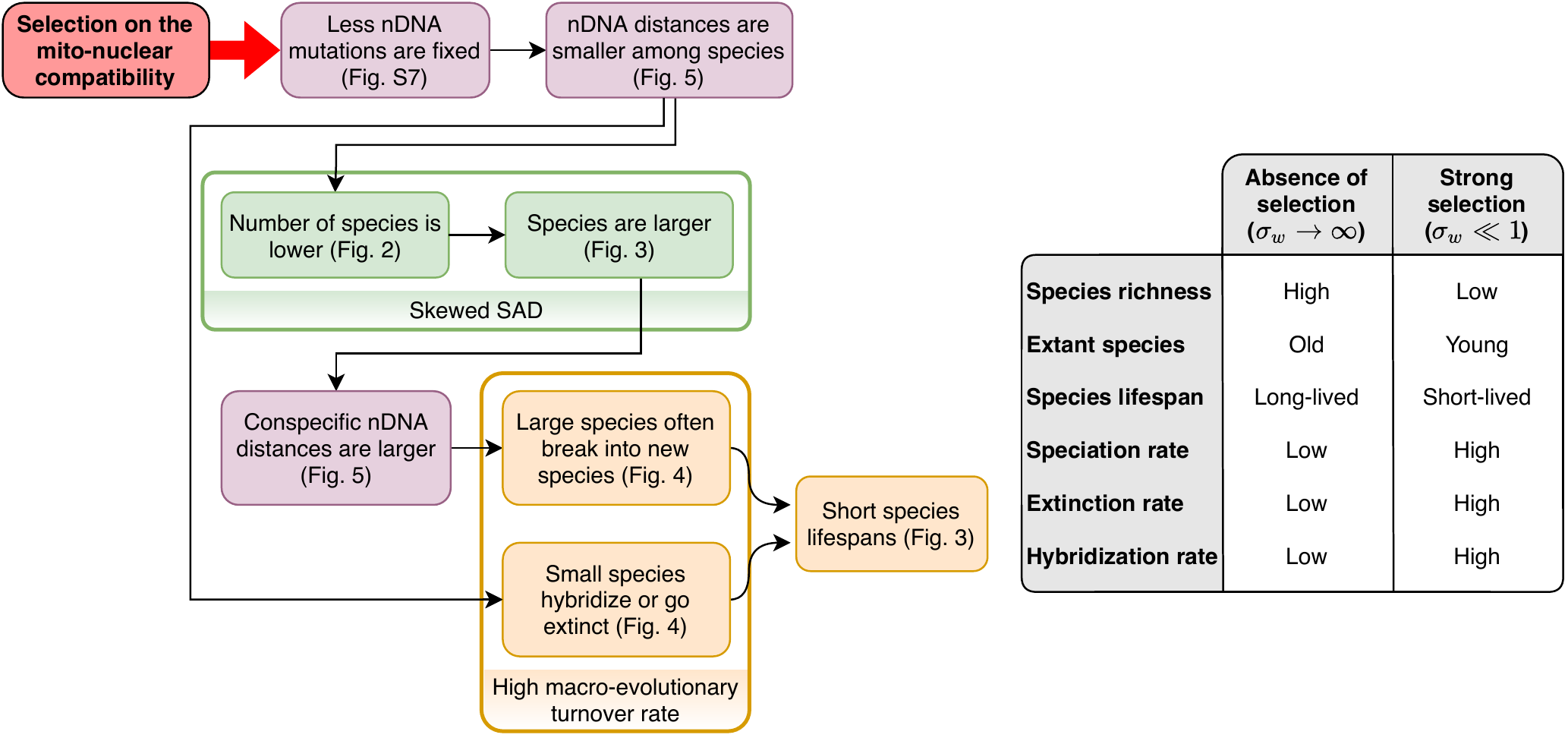}
\end{center}
	\caption{\textbf{From selection at the individual level to ecological and macro-evolutionary patterns.} The flow chart (left) summarizes the cascade of events connecting selection on mito-nuclear compatibility within individuals to its effects on populations, species, and evolutionary turnover. Colors represent the scale of effects: genetic (purple), ecological (green), and evolutionary (orange). The table (right) compares the equilibrium eco-evolutionary patterns in populations under weak versus strong selection. These predictions are qualitatively similar to  empirical ecological and evolutionary patterns in tropical (weak selection) versus temperate zones (strong selection).}
\label{fig:chart}
\end{figure}

A chain of effects connects selection on mito-nuclear genetic compatibility within an individual to skewed species abundance distributions in a community, shorter species lifespans, and elevated macro-evolutionary turnover (chart at Fig. \ref{fig:chart}). Strong selection promotes the conservation of the nuclear genetic content in the population because mutations accumulate more slowly. Consequently, the disruption of gene flow is hindered, and the number of species diminished. With a reduced number of species, they are naturally more abundant and support more diversity, favoring evolutionary branching events. Therefore, the speciation rate increases, and abundant species have reduced lifetimes. 
Low-abundant (non-dominant) species are also short-lived 
%due to the shorter interspecies distances, neighbor species could easily incorporate individuals from these small species, causing abundance fluctuations that increase the probability of extinction. 
for being more susceptible to fluctuations in the abundance.
The reduced genetic distances between different species also promotes hybridization, with sister species merging more easily after speciation. As a global result, communities under mito-nuclear selection were characterized by a low number of species with larger abundances, short species lifespans and high rates of speciation, extinction, and hybridization, i.e., high turnover rate. 
 
\section{Discussion}
 
We have investigated how micro-evolutionary processes impact ecological and macro-evolutionary patterns in a model with selection for mito-nuclear compatibility. Speciation, hybridization, and extinction are emergent phenomena in our model, byproducts of changes in allele frequencies in the populations, as well as spatial and genetic restrictions on mating. We have shown that strong mito-nuclear selection at the individual level triggers a succession of effects in populations and species, with low species richness in equilibrium, reduced species lifespans, and high evolutionary turnover (Fig. \ref{fig:chart}). We investigated the mechanisms by which selection induces these effects, and how they are correlated across  scales.

The negative correlation between species diversity and speciation rates predicted by our model has been observed empirically --  in particular, across latitudinal gradients. The number of species in a taxa typically increases towards the tropics while showing reduced speciation rates, compared to temperate zones \cite{Weir2007, Botero2014, Weir2014, Schluter2017, Harvey2020, Rabosky2018}, along with increasing ages of sister species \cite{Weir2007,Schluter2017,Kennedy2014}. The properties of populations evolved under strong mito-nuclear selection in our model are thus broadly similar to empirical populations in temperate zones, whereas model predictions under weak or no mito-nuclear selection are similar to empirical patterns in the tropics (see table in Fig. \ref{fig:chart}). These empirical patterns are likely the result of many different ecological, evolutionary, and biogeographical processes \cite{Hagen2021, Rangel2018,Pontarp2019}, or even neutral processes \cite{Etienne2019}, such that there is no consensus about the causes of the richness gradient \cite{Mittelbach2007,Hillebrand2004, Kennedy2014} or even the homogeneity of patterns \cite{Evans2005} (ecototherms \cite{Schluter2017} versus planktonic foraminifera\cite{AP2006}, for instance, exhibit a positive correlation between richness and speciation rate). %frogs \cite{Pyron2013}
% The most common explanations for biodiversity hot spots base on more rapid diversification rates \cite{Emerson2005a,Mittelbach2007}, higher carrying capacities\cite{Pontarp2017,Rabosky2015}, or greater time to accumulate species\cite{Wiens2015}, and some hypothesis even combine these factors.
We suggest that selection for mito-nuclear compatibility is a possible mechanism that connects environmental harshness, population genetics and species trajectories. 
In this way, our study helps to link macro-evolutionary patterns of diversification and species turnover, from first principles of mating and fitness operating within a population \cite{Alencar2021}. 

% compared to a previous (neutral) null-model on metacommunity level \cite{Etienne2019}, we have micro-evolutionary process - we are similar cause do not include eco-evolutionary feedbacks (but does still have a ``internal'' selection, within each individual, for mito-nuclear compatibility).\\

Temperate zones are generally regarded as harsh environments with low resource availability \cite{Botero2014,Weir2014,Cutter2016}, in which  metabolic function is critical \cite{Dowling2007}. We therefore assume that selection on mito-nuclear co-adaptation is stronger in such zones. 
% -- \cite{Dowling2007} already implied this association between environmental conditions and selection for co-adapted mito-nuclear genotypes.
The empirical correlation between environmental harshness and the differentiation within species \cite{Weir2014,Botero2014} suggests that speciation is frequent in those regions, which is supported by the observation of young species ages \cite{Smith2017}. Cutter and Gray \cite{Cutter2016} scrutinized how harsh environments promote the recurrent fragmentation of species in space, which may sequentially develop reproductive barriers \cite{Rosindell2010,Li2018}. Those incipient species evolve through local adaptation and ecological speciation, and are therefore more prone to hybridization. Our model produces similar patterns in a process driven by stabilizing (purifying) selection, in contrast to the mechanism of ecological speciation (based on divergent selection) -- which shows that the same macro-evolutionary pattern can result from distinct population processes.

The genetic patterns in our simulated populations reflect the primary causes of variation in diversification rates.
A negative correlation between speciation rates and species richness based on a spatially explicit genetic model was also obtained in \cite{Melian2012}, as a result of sexual reproduction (independent of selection strength). 
In our model, by contrast, strong mito-nuclear selection  reduces the nuclear genetic distance between species (Fig. \ref{fig:dist}a), as the maintenance of mito-nuclear compatibility slows the fixation of mutations for both nDNA and mtDNA (Fig. \ref{fig:subs}). Substitutions occur more slowly, delaying and prolonging radiation and promoting the conservation of genetic content in the population. Since gene flow is higher, there are fewer species that are each more abundant and contain greater intraspecific genetic diversity (Fig. \ref{fig:dist}b,c). 
%In a community under weak selection, by contrast, mutations fix at higher rates resulting a greater number of species each with smaller abundances \cite{Verdu2007}. 
Those abundant species branch more frequently, shortening their lifetimes and resulting in a high speciation rate despite the lower substitution rate of the nuclear alleles. 
Notably, the average rate of substitutions was not positively correlated with the speciation rate\cite{Freeman2022}, which has previously been proposed to be a driver of biodiversity (revised by \cite{Mittelbach2007}).
The large intra-species genetic diversity in communities with low richness in our model are similar to wider niches \cite{Vellend2005, Pontarp2017} and are consistent with predictions of the niche hypothesis (which proposes that ecological opportunity is greater where diversity is low, leading to faster evolution at high latitudes\cite{Freeman2022}).
%which we believe is justified by the convergence of mechanisms involved, albeit with a different motivation.

Although our model reproduces the broad empirical signatures of mito-nuclear coevolution across the genomes \cite{Zhang2013,Piccinini2021}, it does not produce strong signatures of positive selection associated with the compensatory evolution of nDNA in response to mutations in mtDNA \cite{Barreto2018,Hill2020}. This is because we have simulated an adaptive radiation starting from perfectly coordinated mtDNA and nDNA, so that the majority of selection on interacting mito-nuclear sites is purifying selection. 

In contrast to rare species that are more extinction-prone, abundant species are expected to be old and long-lived, i.e.~more robust to the ecological drift \cite {Volkov2003,Chisholm2014}. However, we found that strong mito-nuclear selection reduced species' lifespans even though they were abundant (Fig. \ref{fig:sad}). 
We conclude that, for abundant species, the intrinsic robustness to the ecological drift was opposed by purifying selection against mito-nuclear incompatibilities that rapidly decreased individual fitness as mutations accumulated.
% (corroborated by the test of similar number of species without selection in \ref{SM8}). 
Frequent speciation by species' fission was responsible for the death of abundant species under strong selection, which explains the relationship between their short lifetime and the corresponding high speciation rate (also suggested by \cite{Ricklefs2003}). 

Genetic divergence between species is generally greater in the tropics than temperate regions, which is usually explained by the long age of tropical species \cite{Schluter2017}. Likewise, intraspecific divergences are generally higher in temperate zones, associated with recent speciation rates \cite{Harvey2017,Araujo2013,Smith2017,Botero2014,Weir2014}. Our results suggest that the process of accumulating intraspecific genetic diversity may be the cause, rather than the consequence, of a species being older or younger. 

% Nevertheless, we believe that exploring the model further can elucidate under what conditions these compensatory mutations can be observed (different mutation regimes, variable selection strength, or incompatible initial genomes, for example), so much so that we suggest it as a platform to study the influence of mito-nuclear selection in speciation driven by my microevolutionary process.

\section{Methods}

\subsection{Model}

We use an individual-based model of reproduction and eventual speciation adapted from \cite{aguiar_global_2009, Costa2018}, following the methodology of previous work \cite{Princepe2021}. We consider a spatially distributed population evolving in a homogeneous environment, in the absence of geographic barriers, and under the influence of neutral micro-evolutionary processes -- mutation and genetic drift. Resources are finite but readily available, so that the population is always close to the carrying capacity $M$\cite{Rabosky2015}. Reproduction occurs through pairwise mating of parents.

Derrida \& Peliti \cite{derrida_evolution_1991} first demonstrated this simple model of sympatric speciation with infinite bi-allelic genomes in the absence of natural selection. For sexual reproduction, the only requirement for speciation to occur is {\it genetic compatibility}, a weak form of assortative mating requiring a threshold genetic distance $G$ between mating pairs \cite{derrida_evolution_1991,aguiar_speciation_2017}. Implementation of finite genomes \cite{Princepe2021} requires also {\it spatial proximity of mating individuals}, which must be separated by a maximum spatial distance $S$, changing from a sympatric to a parapatric mode of speciation \cite{aguiar_speciation_2017}. We call the area of radius $S$ around an individual its {\it mating neighborhood.} These are the minimal conditions that guarantee the evolution of reproductive isolation \cite{Thibert-Plante2013}, and they allow us to introduce selection on mito-nuclear compatibility into the model.
Species emerge from the dynamics as groups of individuals connected by gene flow and separated from all others by the genetic mating restriction $G$. Not all individuals of a species need to be compatible since gene flow can be established through intermediary individuals. 
In other words, species correspond to the components of a network where individuals (nodes) are connected if their nuclear genetic distance is less than or equal to $G$. 
The model predicts patterns of biodiversity, such as species abundance distributions (SADs) and species-area relationships \cite{aguiar_global_2009}, similar to previous neutral models \cite{Hubbell2001,Volkov2003}. 

In order to include mito-nuclear selection, individuals are described by a nuclear DNA, which undergoes recombination and defines genetic compatibility for mating, and by a mitochondrial DNA, inherited from the mother. Both genomes are modeled as binary strings with sizes $B$ and $B_M$ respectively. Sexes are separated and individuals are randomly distributed over a square area of size $L \times L$ (see Fig. \ref{fig:model}a) with periodic boundary conditions (individuals can, by chance, occupy the same lattice site). Generations do not overlap and the population is fully replaced by the offspring. Each individual has a probability $P$ to reproduce and $Q=1-P$ of dying without leaving offspring. For reproduction, a mating partner is randomly chosen among genetically compatible individuals in its mating neighborhood. The nuclear genetic distance $D_n$ between individuals is defined as the Hamming distance between their nuclear genomes, and compatibility requires $D_n \leq G$. If a genetically compatible mate is found within the mating neighborhood of the `focal' individual, an offspring is produced with locus by locus recombination of the nuclear genomes, followed by mutation with probability $\mu$ per locus. The mtDNA is inherited from the mother, with mutations with probability $\mu_M$ (Fig. \ref{fig:model}a). The offspring is placed at the site of the focal individual or, with diffusion probability $D$, in one of $S_D$ randomly chosen nearest neighbor sites. Note that individuals can generate more than one offspring: when selected as mating partner by their neighbors or when playing the role of focal individual. If the focal individual cannot find a compatible mate in its mating neighborhood, it expands his search area to radius $S+1$ and then to $S+2$. If no compatible mate is found in the extended area, or if the individual dies without reproducing (probability $Q$), another individual is randomly selected within its original mating neighborhood to reproduce in its place, keeping the population size constant. On rare occasions, however, due to population fluctuations, the mating neighborhood might be empty and no replacement can be found. In these cases, mating does not occur and the population size decreases by one. In order to restore the population to its carrying capacity $M$, we allow individuals to have two offspring in the next generation if their mating neighborhoods have densities below $60\%$ of the expected average value $M/L^2$. This procedure introduces small fluctuations in population size.

We compare the neutral process, when mito-nuclear selection is absent, with the scenario where mito-nuclear compatibility promotes fitness differences in the population, testing different levels of selective pressure while keeping all other parameters fixed. The mito-nuclear distance $d$ measures the compatibility between the mtDNA and the corresponding loci of nDNA, assigned to each individual, ranging from 0 (totally unmatched sequences) to 1 (fully compatible sequences); for completely random sequences, the expected value of $d$ is 0.5. The individual's  fitness $w$ depends on $d$ according to a Gaussian function with width $\sigma_w$, which quantifies the selective strength of the mito-nuclear interaction (Fig. \ref{fig:model}b). The parameter $\sigma_w$ is related to the population average $\langle d \rangle$ in equilibrium (see Fig. \ref{fig:dmed}), such as the absence of selection leads to $\langle d \rangle \approx 0.5 $ and increasing the selection diminishes $\langle d \rangle$ towards 0. Selection over mito-nuclear compatibility changes the probability of an individual to reproduce, i.e., it affects its fecundity. The focal individual has its chance for reproduction modified by its fitness as follows: first the fitness of all individuals in the population are computed as $w(i)=e^{-d(i)^2/2\sigma_w^2}$; second, $Q$ for individual $i$ is modified to $Q_w(i)= 2Q(w_{max}-w(i))/(w_{max}-w_{min})$ where $w_{max}$ and $w_{min}$ are the maximum and minimum fitness in the population, respectively. Individuals with $w(i)=w_{max}$ have $Q_w(i)=0$ and probability of reproducing $P_w=1-Q_w=1$. Those with $w(i)=w_{min}$ have $Q_w(i)=2Q$ and $P_w=1-2Q$. Therefore, individuals with low fitness still have a small chance of reproducing. If the individual dies without reproducing, another one is selected in its mating neighborhood, also according to fitness, to reproduce in its place, keeping the total population constant. Finally, a mating partner is selected from the compatible individuals in the mating neighborhood, with probability proportional to their fitness. 

For the analysis, we considered a standard set of parameters following previous demonstrations of the model that explored the evolution of species under different conditions \cite{aguiar_speciation_2017,Costa2018}. We choose a population size $M$ and a nuclear genome length $B$ that guarantee the onset of radiation and establishment of the equilibrium of the number of species after a few hundred generations. The chosen parameters values were: a population of $M=1,300$ individuals randomly distributed in ($100 \times 100$) lattice with reflecting boundary conditions; the nuclear genome length is $B=1,500$, with mutation rate $\mu=0.00025$, genetic similarity threshold $G=75=0.05B$ and mating range $S=5$. Population density is 0.13, indicating many empty sites in the lattice; the average number of individuals in the mating neighborhood is 10. The mitochondrial genome length is $B_M=500$, with mutation rate $\mu_M = 3 \mu$. We start from an initial condition where all individuals are identical, thus emulating a radiation process. We also fixed $D=0.02$ and $S_D=20$. In this work, we simulate the evolutionary process for $T=2,000$ generations. We fixed $Q=0.37\approx e^{-1}$ as the neutral probability of dying without leaving offspring (equal to the chance of not selecting one particular individual in $M$ trials with replacement \cite{aguiar_speciation_2017}). We have run 50 independent realizations of each scenario and present the population-averaged or accumulated measurements.

\subsection{Processes of diversification}

Throughout the evolutionary dynamics, recombination and mutation change allele frequencies in the population and, combined with the genetic and spatial restrictions for mating, reproductively isolated species can emerge. New species are born from an abundant one that eventually divides into smaller clusters with disrupted gene flow. Due to the model's stochasticity, species abundances and species richness fluctuate randomly  (\textit{ecological drift}). The number of individuals of a species can progressively diminish until no offspring is left to the next generation, due to lack of compatible partners in the mating neighborhood or by chance, and the species is extinct. Also, genetically close species can merge by the reestablishment of the gene flow due to mutations, hybridizing. Therefore we identify three classes of events in which a species can disappear when registering the evolutionary history of the populations: \textit{speciation or branching}: when an abundant species breaks into two or more new species. The mother species is considered dead and new species are born, adding to the number of living species; \textit{extinction}: when the number of individuals of the species progressively diminishes until it reaches a null abundance; and \textit{hybridization}: when the descendants of individuals from two or more different species at $T-1$ belong to the same species at the generation $T$, reestablishing the gene flow. In this case we adopted the protocol of maintaining the largest species and consider the smaller species dead, absorbed by the largest one.

Branching events increase species richness, while extinction and hybridization decrease the number of extant species, reducing biodiversity. Hybridization is an emergent event in our model and only cause loss of diversity. In nature, however, hybridization may increase species richness inducing the evolution of reproductive barriers \cite{Seehausen2004}. Here we emulated the loss of diversity due to hybridization from lineage fusion, speciation reversal \cite{Kearns2018}, and hybrid breakdown  \cite{Barreto2013}. We included in this category events of failed speciation attempts (when two or more recent sister species merge back) and events of reticulate evolution (when there are two speciation events in a row and a species fuses with the one with the most distant ancestor)

\subsection{Species abundances distributions}

In our simulations the number of species in equilibrium ($T \geq 500$) ranged from about 20 in the non-interacting case to about 5 for strong selection (Fig. \ref{fig:acc}b). To generate abundance distribution plots with statistical significance, we have run 50 independent simulations for each set of parameters and accumulated the results. This increased the number of species in the histograms to about 1,000 for the non-interacting case and 100 for the case of strong selection. Section \ref{SM2} of Supplementary Information justifies the validity of this procedure.

To generate the species' abundances histograms, data were binned in abundances classes (log2). There are several methods for choosing the bins' boundaries, and we followed the method adopted by \cite{Hubbell2001}: bins were built by using the powers of two as the center of bins. The boundaries then come as $2^{n-1/2}$ and $2^{n+1/2}$. Thus,  bin=0 stores species with 1 individual, bin=1 counts species with 2 individuals, bin=2 species with 3, 4, and 5 individuals, and so on.

\subsection*{Data availability}

\noindent There are no empirical data associated with this study.

\subsection*{Code availability}

\noindent All simulations are run in Fortran. All the codes for simulations and scripts for data treatment in Python are available in the GitHub repository at https://github.com/deborapr/mito-nuclear-speciation.

\subsection*{Acknowledgment}
% \noindent DP was supported by the São Paulo Research Foundation (FAPESP), grants \#2018/11187-8, \#2019/24449-3 and \#2016/01343-7 (ICTP-SAIFR).
\noindent This work was partly supported by the São Paulo Research Foundation (FAPESP), grants \#2018/11187-8 (DP), \#2019/24449-3 (DP), \#2019/20271-5 (MAMA) and \#2016/01343-7 (ICTP-SAIFR). MAMA was supported by Conselho Nacional de Pesquisas Científicas (CNPq), grant \#301082/2019-7.

\bibliographystyle{naturemag}
\bibliography{references}

\makeatletter\@input{file1.tex}\makeatother

\end{document}

\doublespacing 
	\selectlanguage{english}
	\sloppy
	\begin{center}
	{\Large Supplementary Information for \\\textbf{Mito-nuclear selection induces a trade-off between \\ species ecological dominance and evolutionary lifespan}}\\
	{\large {\textit{Débora Princepe, Marcus A. M. de Aguiar and Joshua B. Plotkin}}}

\end{center}

	\tableofcontents
	
	\pagebreak

\section{Diversification rates}
\label{SM1}

In this section we show how the diversification rates -- speciation, hybridization and extinction -- depend on the selection strength at equilibrium and how they behave along the evolutionary process (radiation and equilibrium).

\begin{figure}[!h]
\begin{center}
    \includegraphics[width=0.4\linewidth]{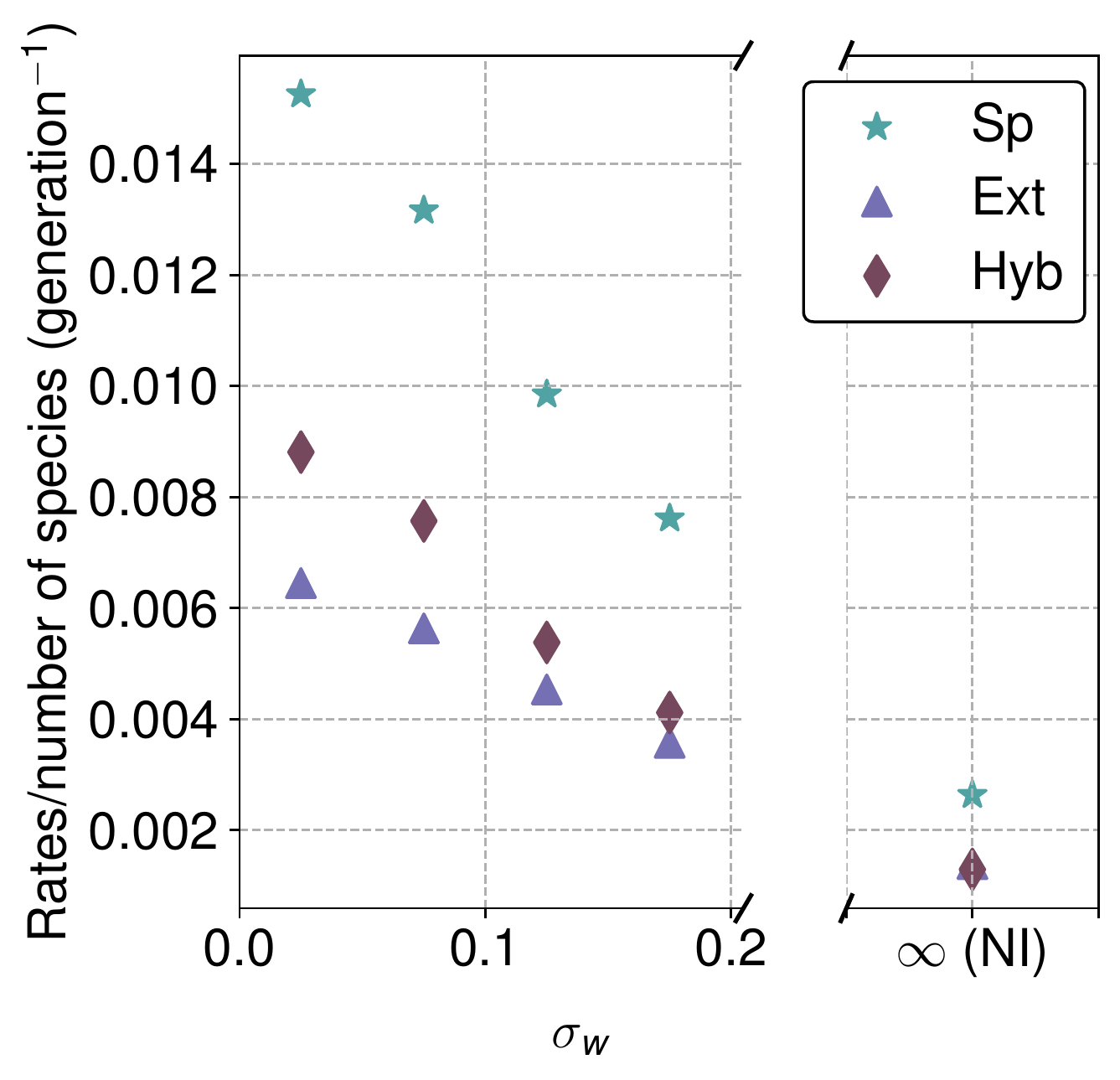}
\end{center}
	\caption{Diversification rates per species in the equilibrium: speciation (Sp), extinction (Ext) and hybridization (Hyb).}
	\label{fig:ratio}
\end{figure}

\begin{figure}[!h]
\begin{center}
    \includegraphics[width=0.5\linewidth]{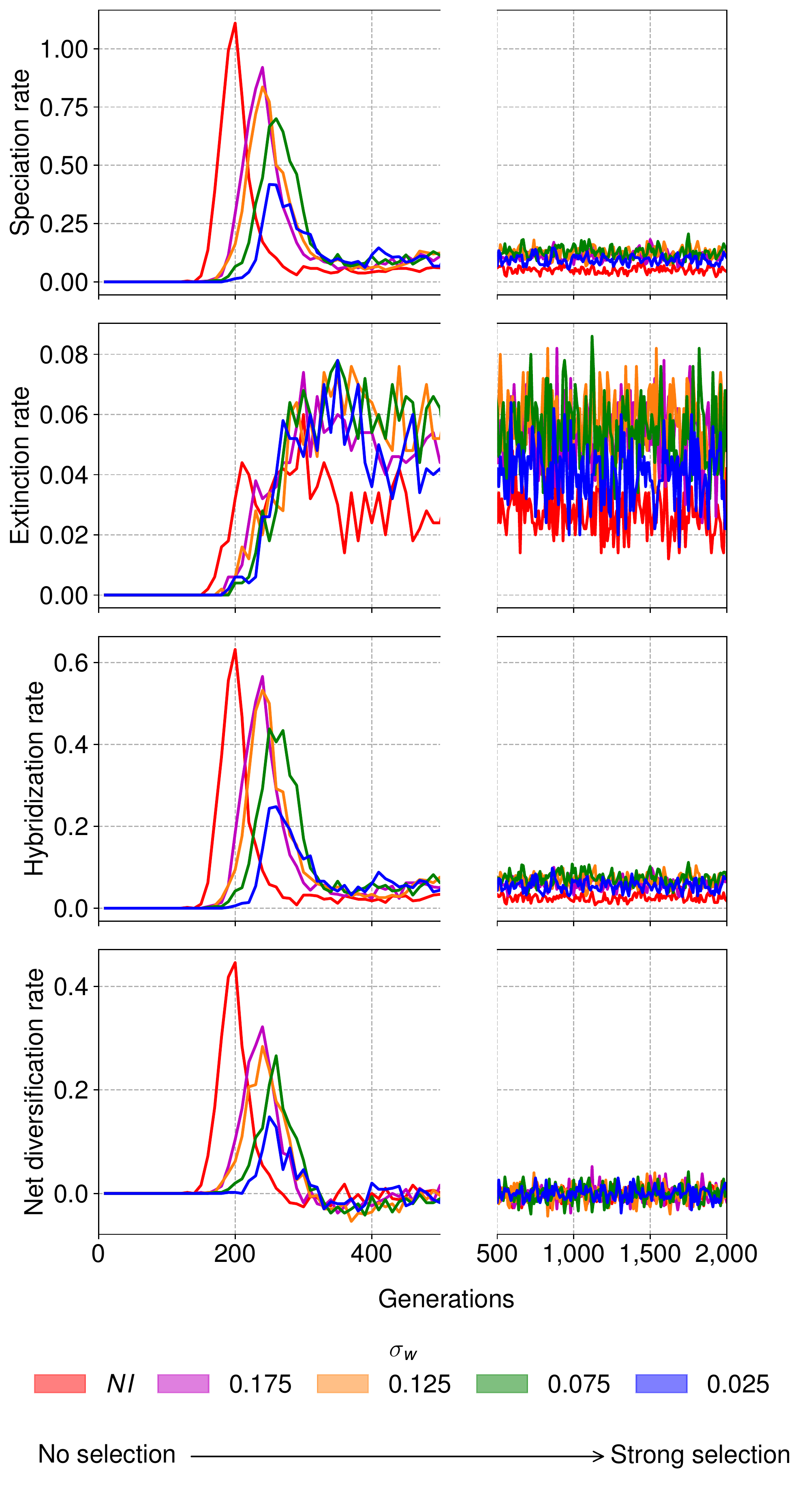}
\end{center}
	\caption{Diversification rates over time. Events were accumulated over time intervals $\Delta t=10$ and the results divided by $\Delta t$. }
	\label{fig:div_rates}
\end{figure}

Figure \ref{fig:ratio} shows the diversification rates per species in the equilibrium regime for different selection strengths $\sigma_w$ and in the absence of selection (non-interacting case, $NI$) (analogous to the rate of speciation per lineage, as usually shown for comparison of global patterns \cite{Schluter2017}). We note that the number of species in equilibrium varies considerably with $\sigma_w$ (Fig. 2b of the main text). Figure \ref{fig:div_rates} show the diversification rates along the evolutionary time (instantaneous rates evaluated for time bins $\Delta t =10$ generations), displaying peaks during the radiation and reaching a steady state afterwards. The net diversification rate, result from speciation minus extinction and hybridization, peaks at the radiation and it is null in the equilibrium, when species richness reaches a plateau.

\clearpage
\pagebreak
\newpage

\section{Scalability of the community size}
\label{SM2}

\begin{figure}[h]
\begin{center}
        \includegraphics[width=0.9\linewidth]{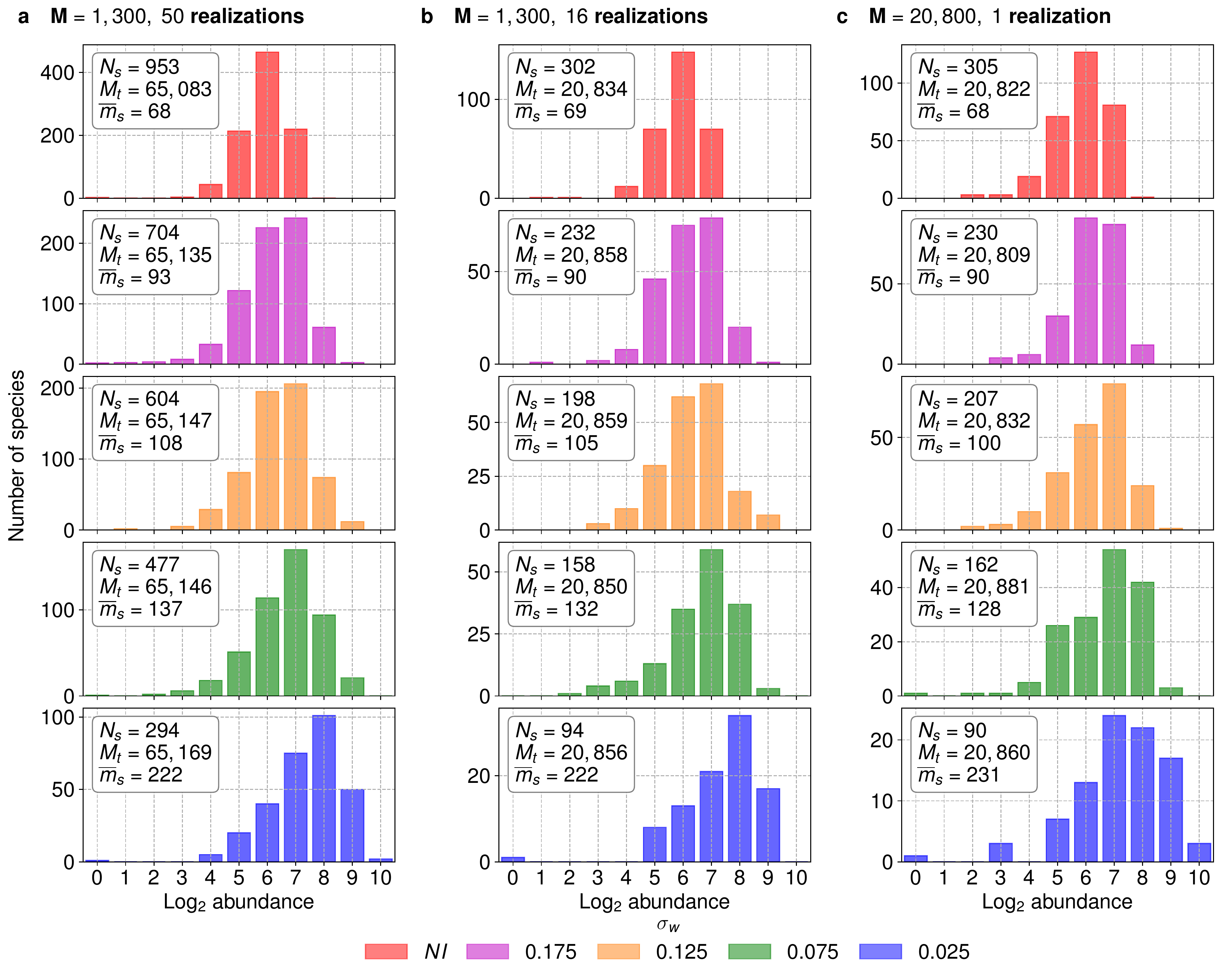}
\end{center}
    \caption{Species abundances distributions accumulated over 50 simulations with 1,300 individuals (a); over 16 simulations with 1,300 individuals (b); and single simulation with 20,800 individuals (c). $N_s$ is the total number of species, $M_t$ is the total number of individuals and $\bar{m}_s$ is the mean species' abundance over all simulations.}
	\label{fig:sad-sample}
\end{figure}

Our simulations resulted in a small number of species in equilibrium. To obtain good statistics for the species abundances distributions (SAD), we have accumulated the results of several independent simulations. We validate this methodology by comparing the resulting distributions for each value of $\sigma_w$ with a single simulation of a large population of $M=20,800$ individuals keeping the demographic density constant, i.e., expanding the area to a $400 \times 400$ lattice. Figure \ref{fig:sad-sample} shows the accumulated result over 50 simulations, comprising $\sim65,000$ individuals (Fig. \ref{fig:sad-sample}a),  accumulated results over 16 simulations with a total of $\sim20,800$ individuals (Fig. \ref{fig:sad-sample}b) and single simulation with 20,800 individuals (Fig. \ref{fig:sad-sample}c). Variation of the number of individuals are due to stochastic fluctuations in mating choice during the simulations. We observed that the SADs were equivalent regarding the number of species sampled and the average species abundance. 

\clearpage
\pagebreak
\newpage

\section{SADs at intermediary times}
\label{SM3}

\begin{figure}[h]
\begin{center}
    \includegraphics[width=1\linewidth]{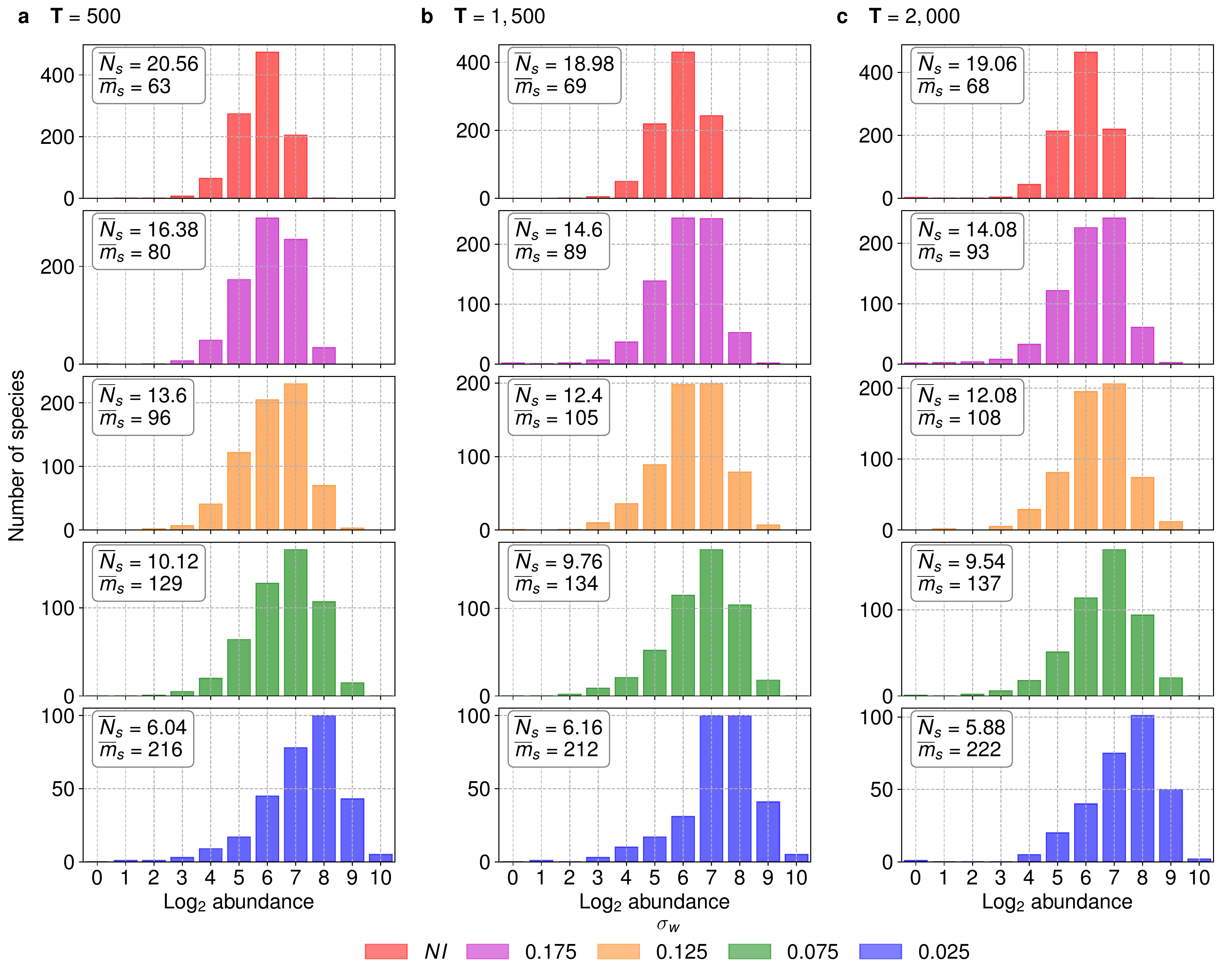}
\end{center}
    \caption{Abundance distributions computed at three different times after the radiation period. $\bar{N}_s$ is the mean number of species and $\bar{m}_s$ is the mean species' abundance. Although small fluctuations are observed, the overall shape does not change significantly over time.}
	\label{fig:sad-time}
\end{figure}

Here we show that, after equilibration at $T\approx 500$ generations, the abundances distributions do not change significantly, reaching a steady configuration. Figure \ref{fig:sad-time} shows that, despite small fluctuations due to the stochastic nature of the model, the overall shape of the distributions do not change.

\clearpage
\pagebreak
\newpage

\section{Mean, variance and skewness of the SADs}
\label{SM4}

\begin{figure}[h]
\begin{center}
    \includegraphics[width=0.38\linewidth]{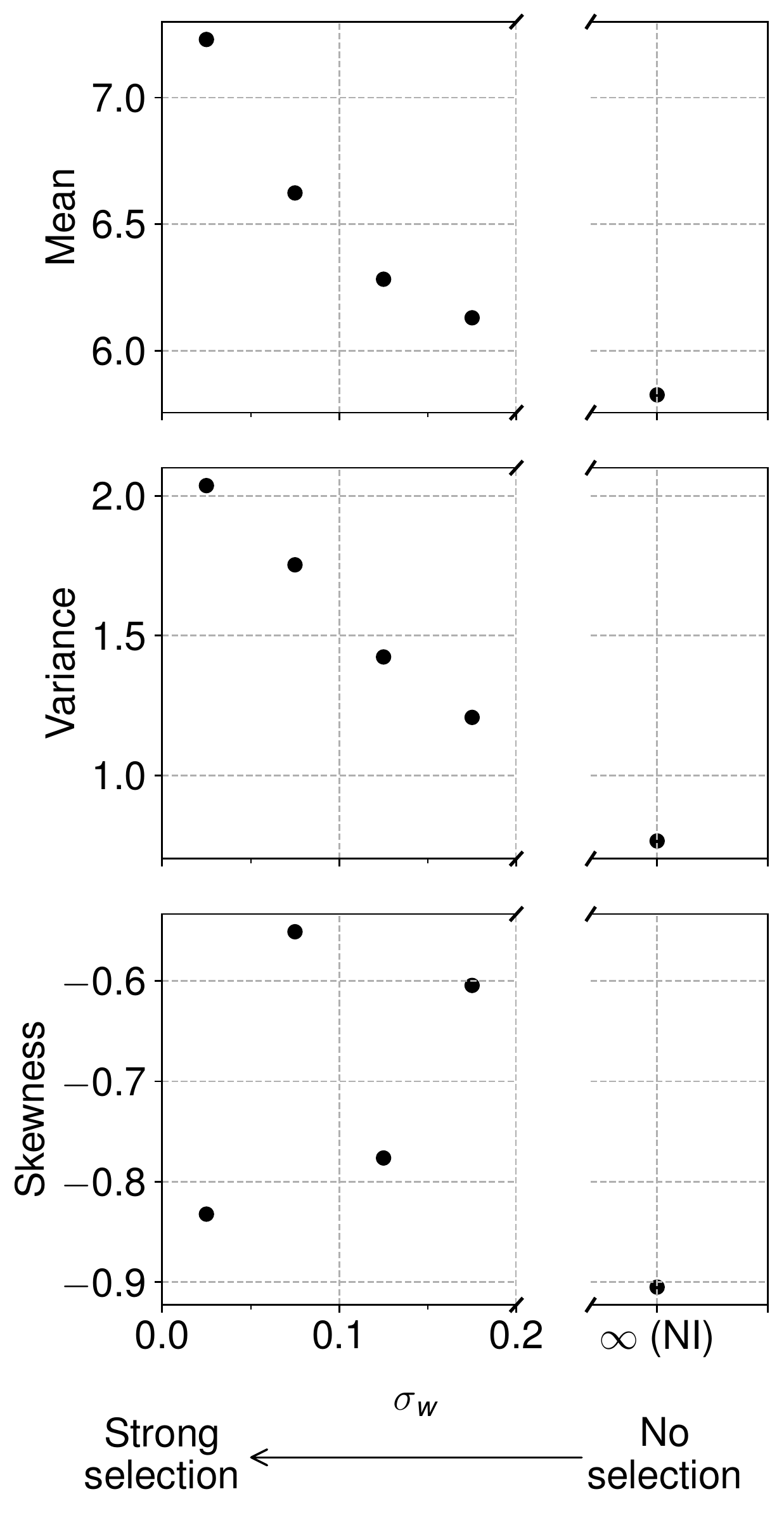}
\end{center}
    \caption{Mean value, variance and skewness of the histograms of Figure 3a.}
	\label{fig:sad-skew}
\end{figure}

The differences between the histograms of Figure 3a (Species abundances distributions at $T=1000$) can be quantified simply by the first, second, and third momentum of the distribution, i.e., the mean value, the variance, and the skewness of the histograms (Fig. \ref{fig:sad-skew}). We considered the center points of the bar diagrams where the abundances were binned in log2 abundances classes, as described in Section 4.3. Therefore, we did not assume any best fit to the data (lognormal, Fisher distribution, etc.) because we aimed only to see the effect of selection on the distribution. Previous work on this model discussed the resulting patterns of species abundances (in the absence of mito-nuclear selection) and compared them with empiric data \cite{aguiar_global_2009}. 

Selection had the role of increase the mean size of species (Fig. \ref{fig:sad-skew}, top) and the distributions also became wider (Fig. \ref{fig:sad-skew}, middle). However, the skewness did not preset a monotonic tendency with increasing selection (Fig. \ref{fig:sad-skew}, bottom), all distributions were left-skewed (skewness with negative value) and the non-interacting model is the most skewed.

\clearpage
\pagebreak
\newpage

\section{Species' ages and time for death}
\label{SM5}

\begin{figure}[h]
\begin{center}
    \includegraphics[width=0.7\linewidth]{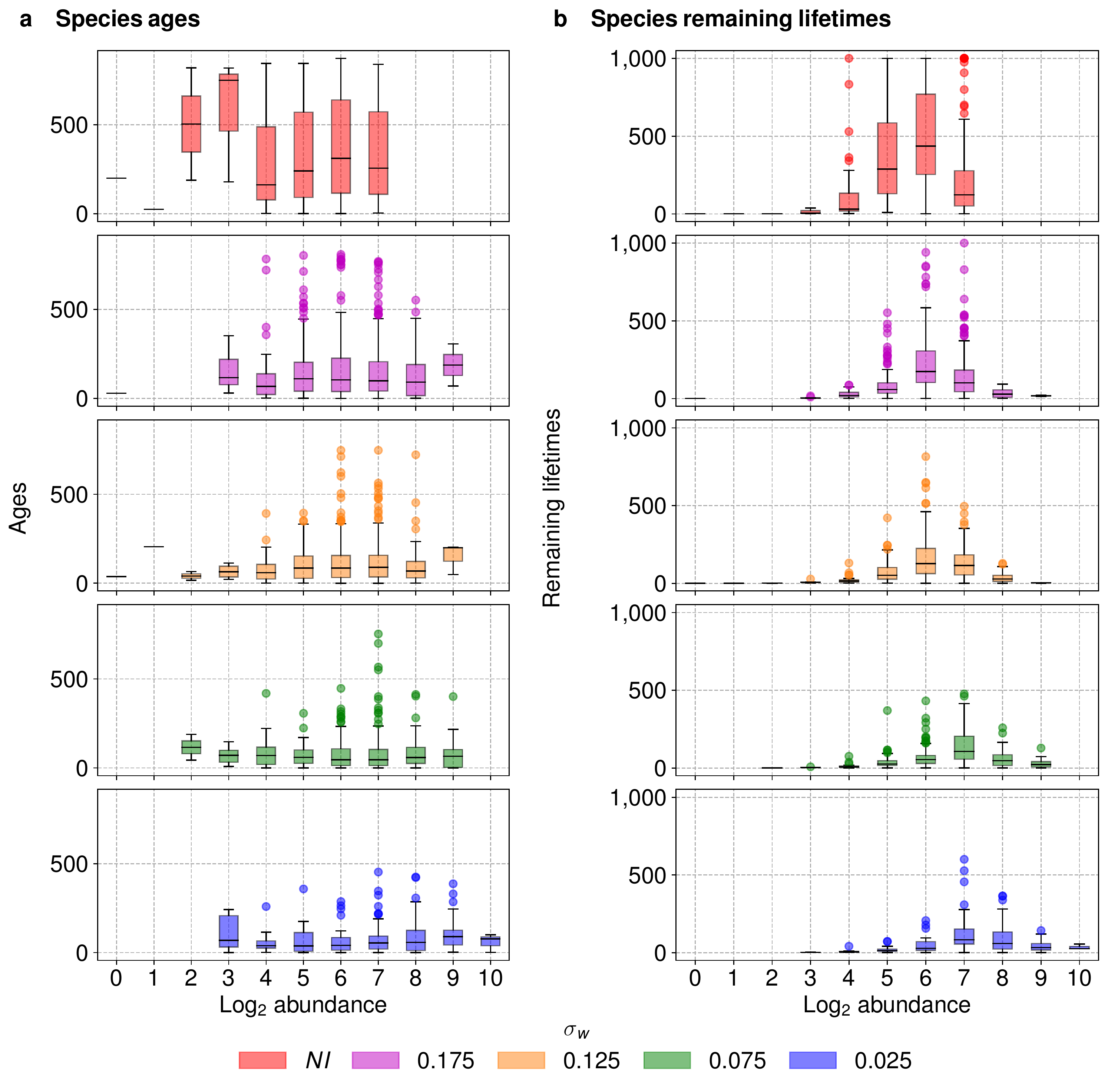}
	\caption{Relation between species abundances and their (a) ages and (b) and remaining lifetimes for species extant at $T=1,000$. }
	\label{fig:remain}
\end{center}
\end{figure}

Here we investigate how species abundances correlate with age at the time of sampling and remaining lifetime. Species were detected at $T=1,000$ and their abundances and ages computed. Figure \ref{fig:remain} shows that species of intermediate sizes are older (Fig. \ref{fig:remain}a) and live longer (Fig. \ref{fig:remain}b) when selection over the mito-nuclear interaction is considered. The non-interacting case shows no significant correlation between lifespans and abundances, except for small species, that tend to go extinct faster.
 
\clearpage
\pagebreak
\newpage

\section{Mitochondrial distances}
\label{SM6}

\begin{figure}[h]
\begin{center}
        \includegraphics[width=0.7\linewidth]{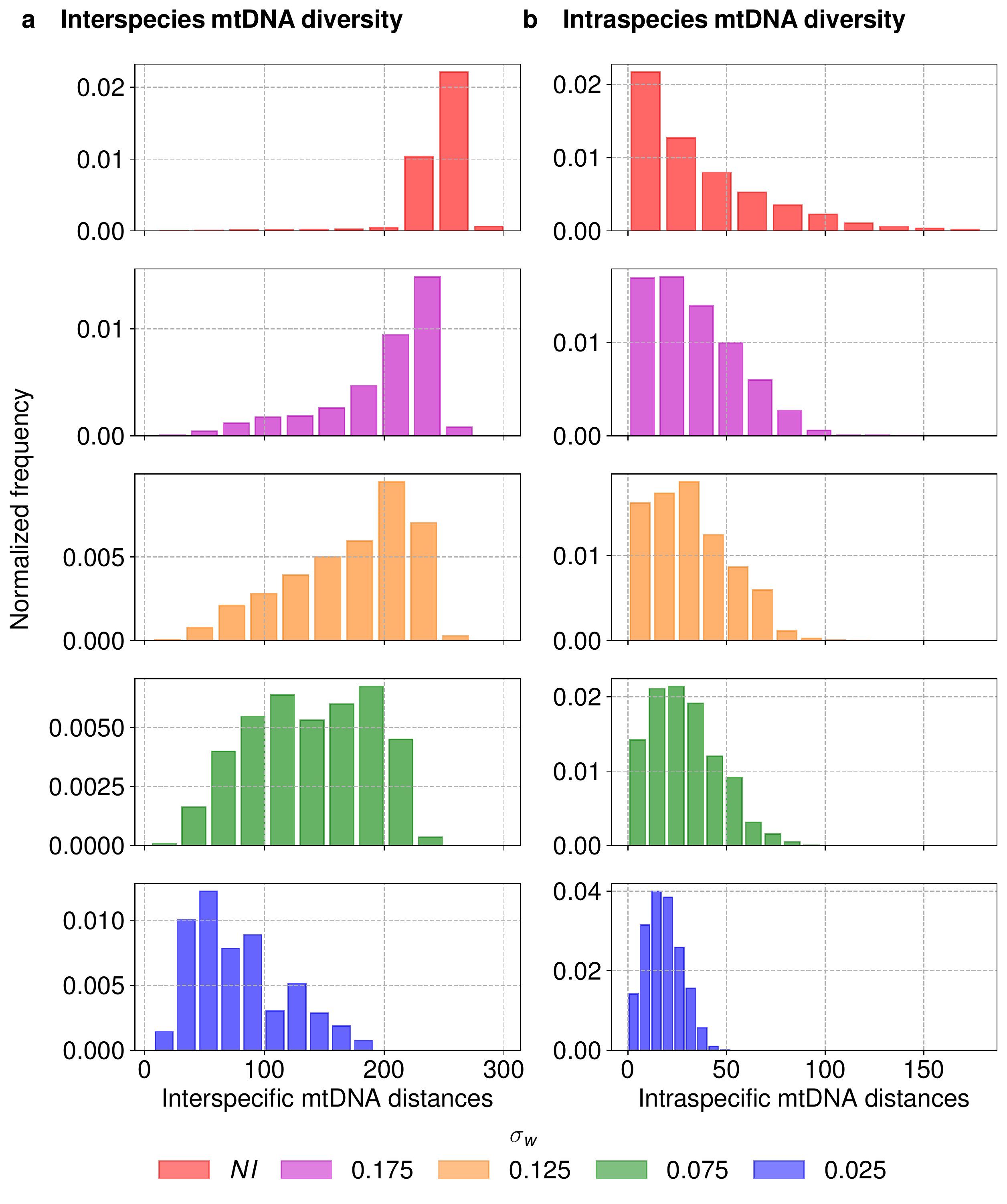}
\end{center}
	\caption{Mitochondrial distances between pair of individuals belonging to different species (a) and from the same species (b). Similarly to the nDNA, mito-nuclear selection reduced the mitochondrial distances between individuals of different species. For individuals of the same species, the mtDNA distances also reduced with selection, in opposite trend to the nDNA.}
	\label{fig:mtdna}
\end{figure}

Figure \ref{fig:mtdna} shows the distribution of mitochondrial distances between (Fig. \ref{fig:mtdna}a) and within (Fig. \ref{fig:mtdna}b) species. Selection over mito-nuclear interaction drastically reduces the interspecies distances, implying in more genetically similar species. Intraspecies distances, on the other hand, change from an exponential distribution with large average distance (non-interacting case) to a unimodal distribution with much smaller average distance. Species, therefore, become more uniform in terms of their mitochondria as selection strength increases. For the mitochondrial DNA, which does not recombine and is not under mating restriction, selection utterly increased similarity within species, spontaneously forming clusters of mitochondria (Fig. \ref{fig:mtdna}b).

% mtDNA is different because: does not recombine and it is not restricted to mating. we showed before that it is highly influenced by spatial mating too. cluster of ndna occurred because mating restriction; clusters in mtdna occurred due to spatial mating and mono-parental inheritance. in this case, selection increased the mtdna similarity within species.

\clearpage
\pagebreak
\newpage

\section{Substitutions rates}
\label{SM7}

In our simulations the mutation probabilities per locus were fixed at $\mu=0.00025$ for the nDNA and $\mu_M=0.00075$ for the mtDNA. Mutation reversals, however, decrease the effective number of mutations and selection filters out `deleterious' mutations that decrease fitness, also decreasing the net rate of substitutions. 

The number of substitutions is computed as the number of loci that changed from the initial state 0 to state 1 at $T=2,000$. For the non-interacting case, the probability that a single mutation occurs and is not reversed after $T$ generations is $\binom{T}{1}\mu(1-\mu)^{T-1}$. The probability that a mutation occurs and is reversed twice is $\binom{T}{3}\mu^3(1-\mu)^{T-3}$. Thus, ${\mu [(1-\mu)^{T-1}+\frac{(T-2)(T-1)}{6}\mu^2(1-\mu)^{T-3}]}$ is an estimate for the average number of substitutions per generation. The expected substitution rates of the nuclear and mitochondrial genomes in the non-interacting case are then $1.5\times 10^{-4}$ and $2.3\times 10^{-4}$ per generation, respectively \cite{Princepe2021}.

\begin{figure}[h]
\begin{center}
        \includegraphics[width=0.6\linewidth]{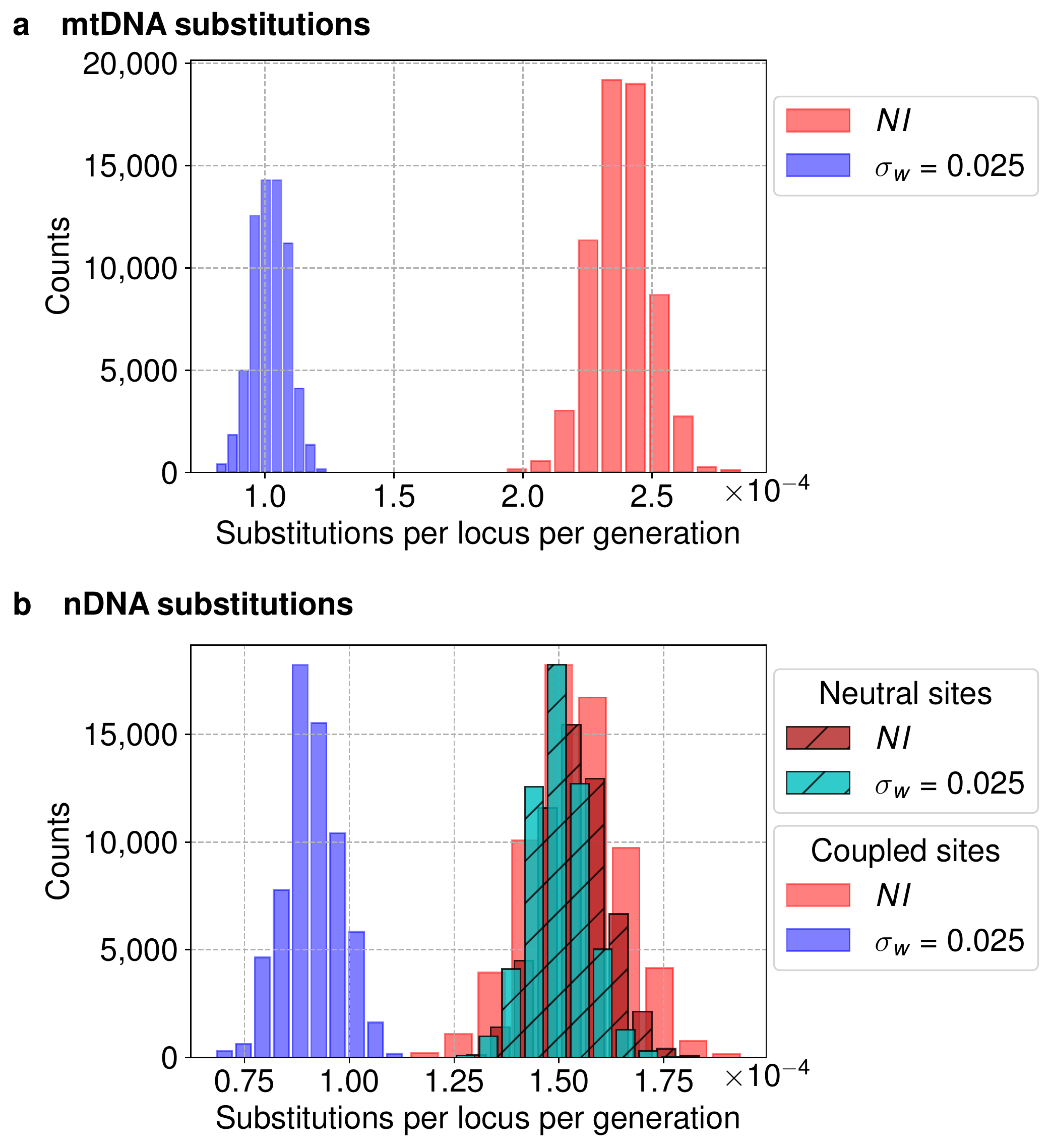}
\end{center}
	\caption{Substitution rates per locus per generation for the mtDNA (a) and for the nDNA (b).}
	\label{fig:subs}
\end{figure}

Figure \ref{fig:subs}a shows the substitution rate per locus in the mtDNA in the non-interacting case (red) and the strongly interacting case (blue). Figure \ref{fig:subs}b shows similar information for the nDNA, separated into coupled sites (first 500 loci) and uncoupled sites (remaining 1,000 loci).

\begin{figure}[h]
\begin{center}
        \includegraphics[width=0.67\linewidth]{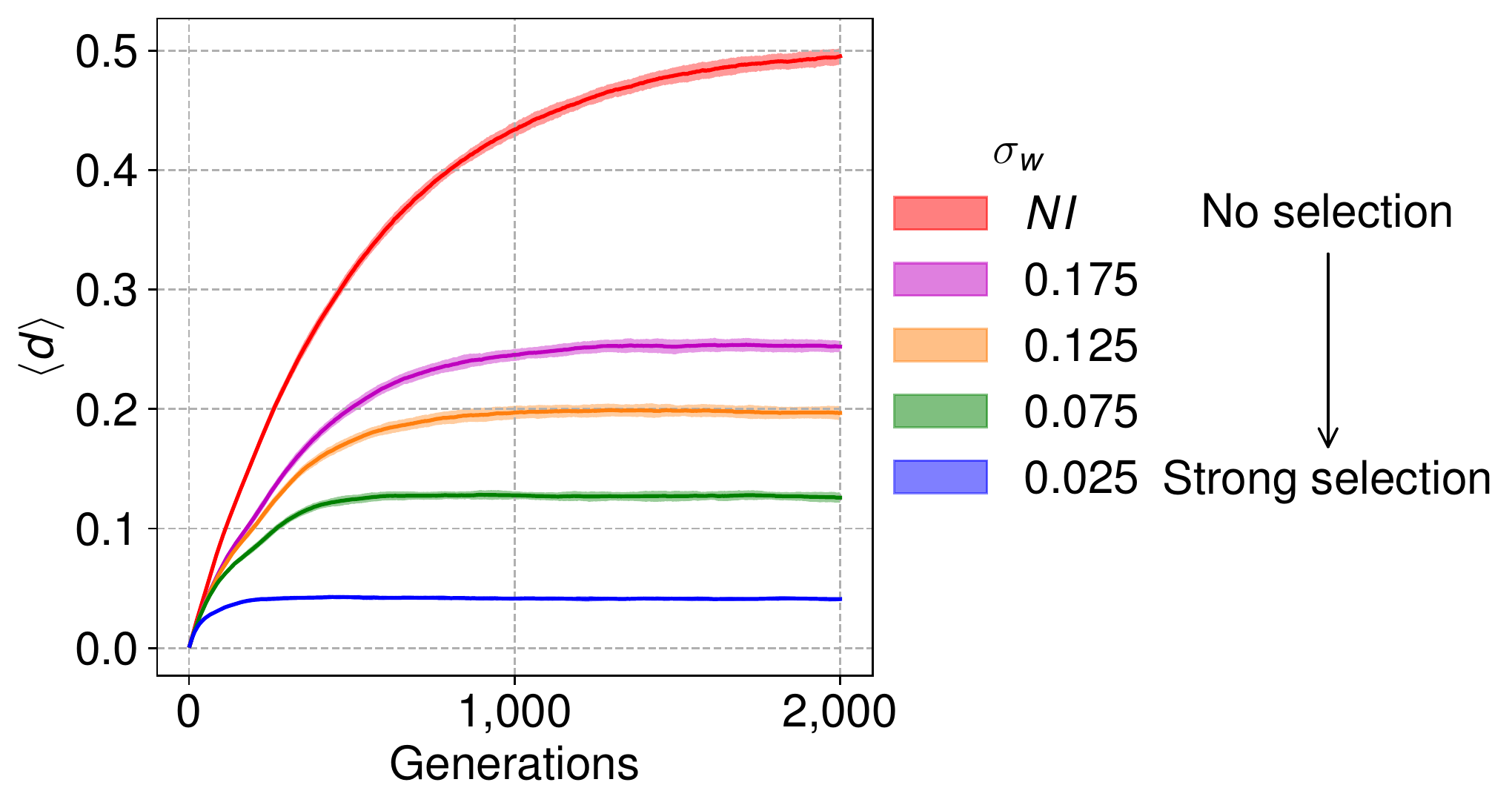}
\end{center}
	\caption{Population averaged mito-nuclear distance $\langle d \rangle$ as function of time. Results are shown for varying strength of the mito-nuclear selection $\sigma_w$ and in the absence of selection (NI). Solid lines represent the ensemble mean over 50 independent realizations, and the shadowed area represents one standard deviation from the mean.}
	\label{fig:dmed}
\end{figure}

The resulting population-averaged mito-nuclear distance, computing the Hamming distance between coupled sites in the nDNA and the respective sites in the mtDNA, is shown in Figure \ref{fig:dmed}. In the absence of selection, the sequences evolve independently, and $\langle d \rangle$ reaches the maximum value of 0.5 (expected value for completely random sequences). When selection is present, the steady-state value of $\langle d \rangle$ is smaller with stronger the selection, i.e., the mito-nuclear compatibility is preserved.

\clearpage
\pagebreak
\newpage

\section{Non-interacting case with reduced number of species}
\label{SM8}

\begin{figure}[!b]
\begin{center}
        \includegraphics[width=0.7\linewidth]{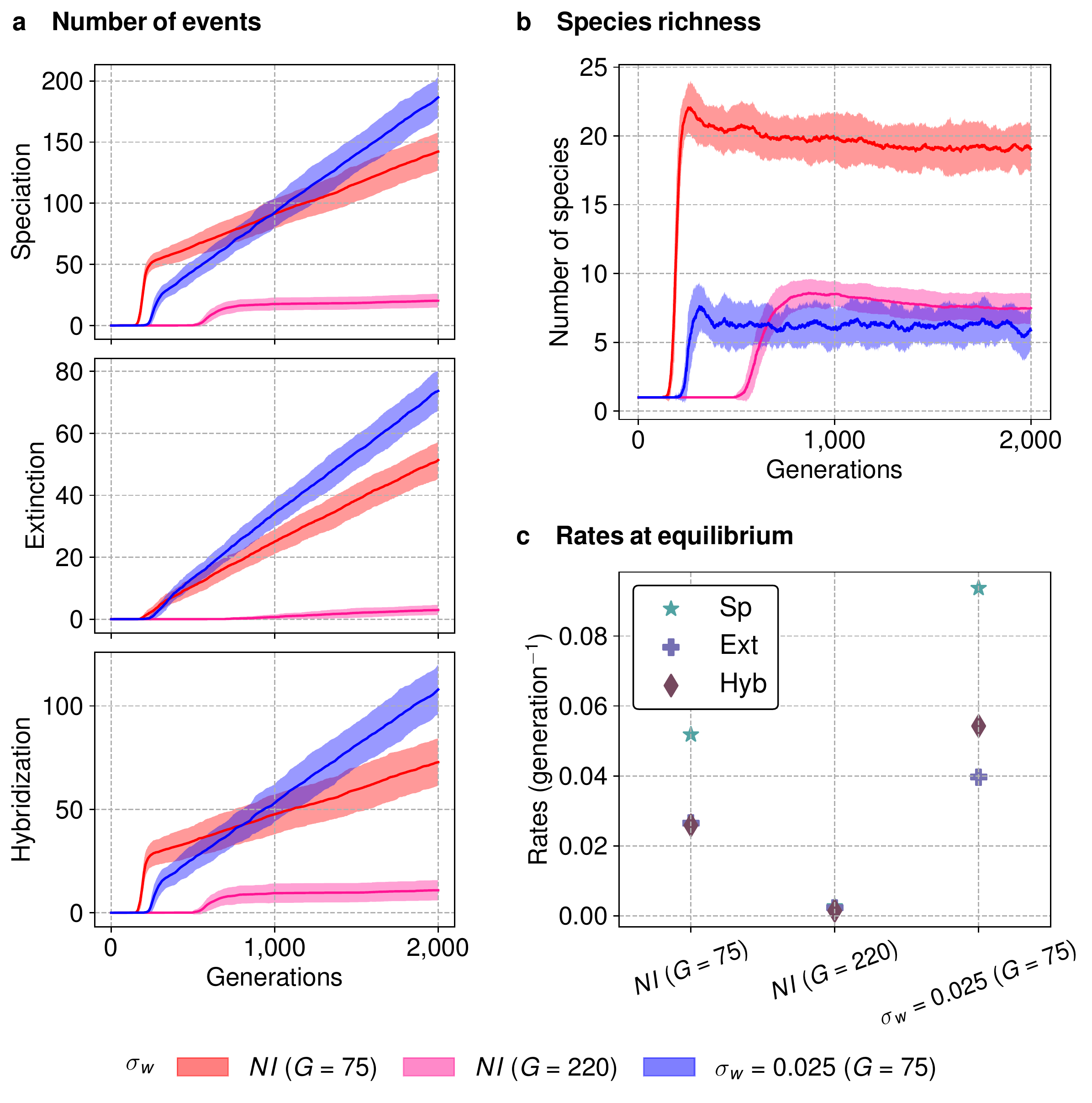}
\end{center}
	\caption{Diversification events and species richness. \textbf{a}, Accumulated number of speciation, extinction, and hybridization events over time. \textbf{b}, Species richness as function of time. \textbf{c}, Rates of speciation (Sp), extinction (Ext), and hybridization (Hyb) obtained from regression of \textbf{a} in equilibrium ($T>1000$). Results are shown in the absence of selection (NI) and with $\sigma_w=0.025$ with mating restriction $G=75$, as presented in the main text, and without selection with $G=220$.}
	\label{fig:g220-nsp}
\end{figure}

In the main text, we compared the evolution of populations under variable selection strength while fixing all other parameters, which resulted in a decrease in the number of species as the intensity of selection increases (Fig. 2b). Here we compare the case of strongest selection ($\sigma_w=0.025$) with a neutral case presenting similar number of species. The purpose of such comparison is to evaluate whether the effects on diversification rates and species' lifespans and fates are simply a consequence of the decrease in species richness. The non-interacting case ($NI$) presented a similar number of species of $\sigma_w=0.025$ when the mating restriction was decreased, setting $G=220$ ($G=0.15B$, in contrast with $G=75=0.05B$ in the main text), as shown in Figure \ref{fig:g220-nsp}b. The radiation process was delayed, and the number of diversification events drastically decreased (Fig. \ref{fig:g220-nsp}a), also reducing the rates at the steady-state (Fig. \ref{fig:g220-nsp}c).  Due to the reduced number of species, they became more abundant, right-shifting the SADs (Fig. \ref{fig:g220-abund}a). Compared to $NI  \; (G=75)$, the mean species' abundance increased (Fig.  \ref{fig:g220-skew}, top), but the distribution was not much wider (Fig.  \ref{fig:g220-skew} middle). More important, species' lifespans were not reduced: comparing $NI \; (G=220)$ and $\sigma_w=0.025$, species in the same abundance class lived longer in the absence of selection than under strong selection (Fig. \ref{fig:g220-abund}b). The remaining lifetimes were also longer (Fig. \ref{fig:g220-ages}b), and, even with the delay in the radiation, species at $T=1000$ were older without selection (Fig. \ref{fig:g220-ages}a). Therefore, the reported reduction of lifespan and diversification rates were due to the selection over mito-nuclear compatibility. %Species' ages and lifetimes were also longer in the absence of selection even with the delay in the radiation process, corroborating our conclusions on the effects of mito-nuclear selection on species' longevity.

\begin{figure}[h]
\begin{center}
        \includegraphics[width=0.7\linewidth]{Figures/supp-g220-abund.pdf}
\end{center}
	\caption{Species abundances distributions (SADs) at $T=1,000$ (a) and respective lifespans (b). $\bar{N}_s$ is the mean number of species over all simulations.}
	\label{fig:g220-abund}
\end{figure}

\begin{figure}[h]
\begin{center}
        \includegraphics[width=0.31\linewidth]{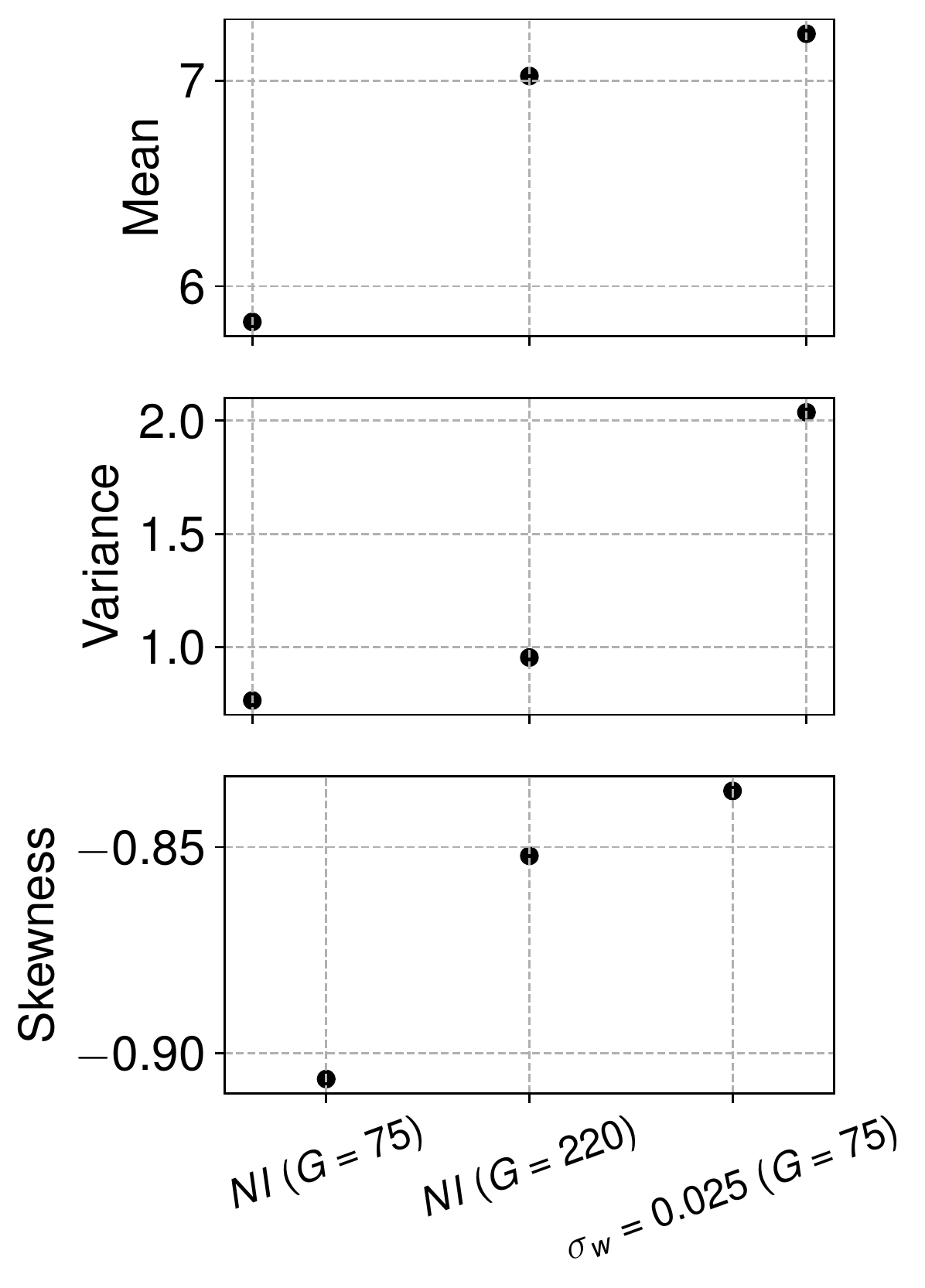}
\end{center}
	 \caption{Mean value, variance and skewness of the histograms of Figure \ref{fig:g220-abund}a.}
	\label{fig:g220-skew}
\end{figure}

\begin{figure}[h]
\begin{center}
        \includegraphics[width=0.7\linewidth]{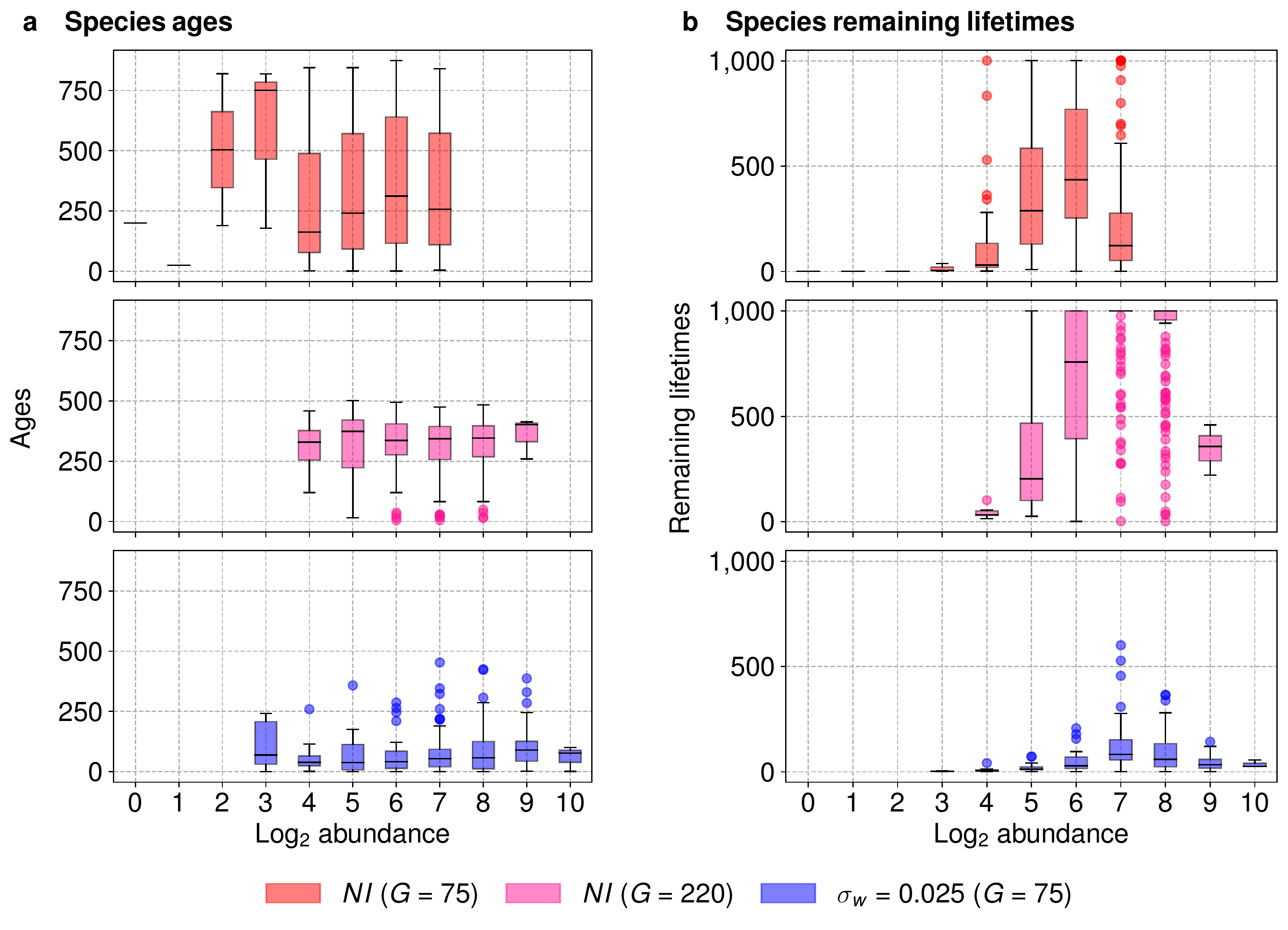}
\end{center}
	\caption{Relation between species abundances and their (a) ages and (b) and remaining lifetimes for species extant at $T=1,000$.}
	\label{fig:g220-ages}
\end{figure}

\begin{figure}[!h]
\begin{center}
        \includegraphics[width=0.4\linewidth]{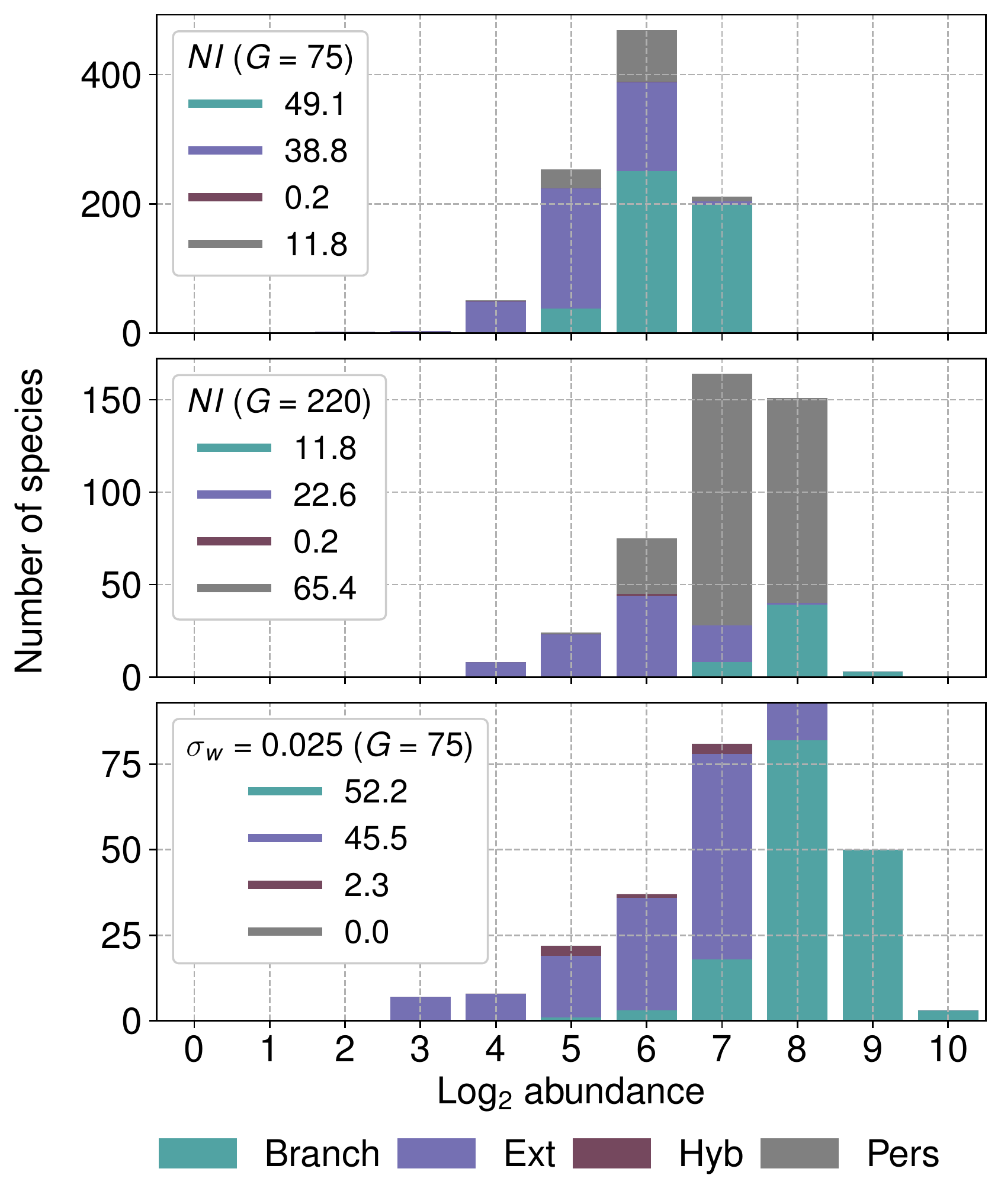}
\end{center}
	\caption{Long-term evolutionary fates of species. The histograms depicts the fate of each species shown in Fig. \ref{fig:g220-abund}a: by branching (Branch), extinction (Ext), or hybridization (Hyb), or persistence through time $T=2000$ (Pers). The legends indicate the percentages of total species that correspond to each event category.}
	\label{fig:g220-events}
\end{figure}

The long-term evolutionary fates also indicate that selection was responsible for increasing the number of branching, hybridization and extinction events. Species on abundances bins 7 and 8, for instance, show very different trajectories in absence or under selection: while species on these classes were most likely to persist until $T =2000$ without selection, they were extincted, hybridized or split into new species when under selection (Fig. \ref{fig:g220-events}).

\begin{figure}[h]
\begin{center}
        \includegraphics[width=0.6\linewidth]{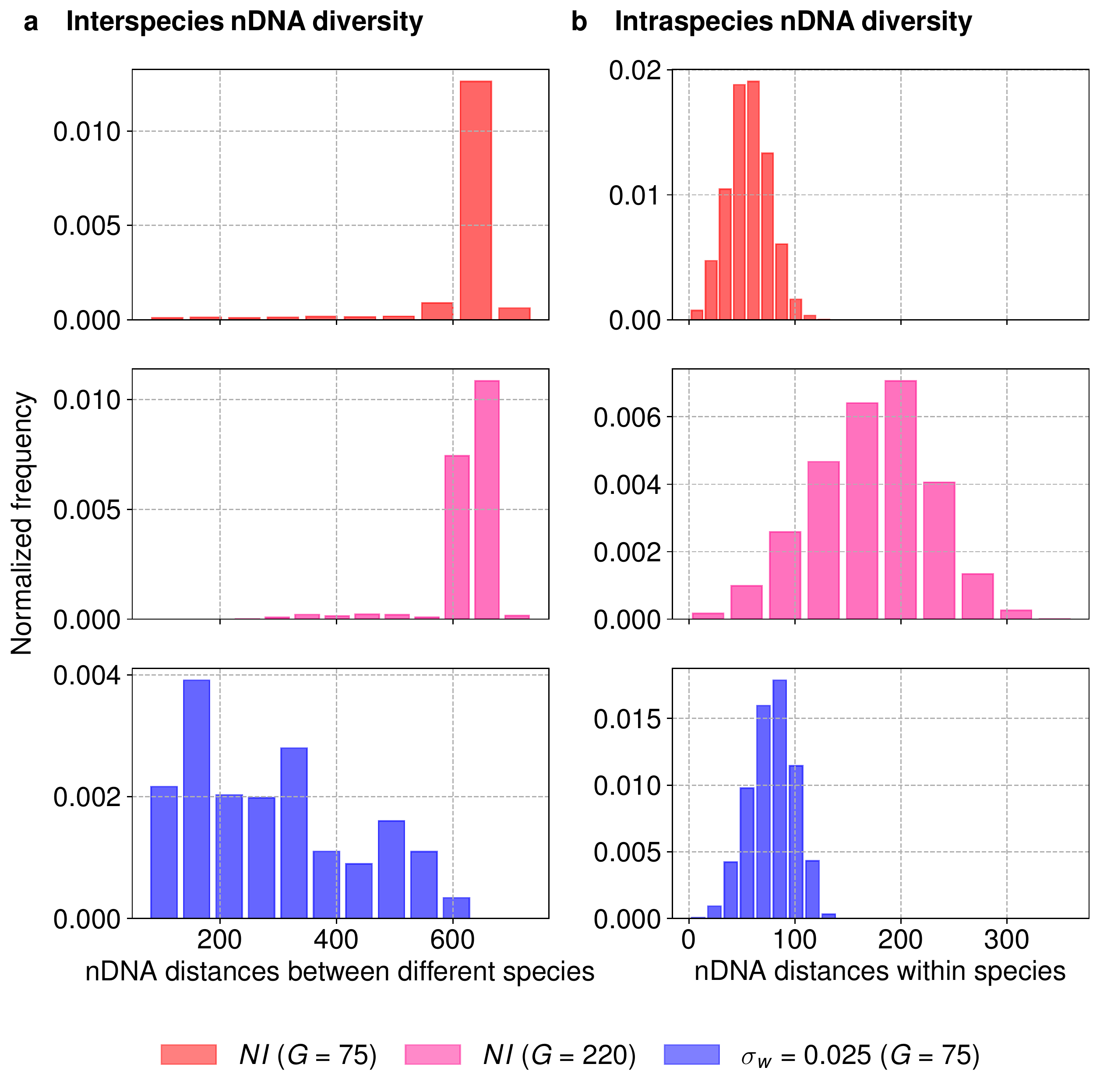}
\end{center}
	\caption{Nuclear distances between pair of individuals belonging to different species (a) and from the same species (b).}
	\label{fig:g220-ndna}
\end{figure}

These evolutionary patterns connect once again to the genetic structure of the population: even though the increase in the genetic threshold $G$ made nDNA distances within species larger (Fig. \ref{fig:g220-ndna}b), the distance between different species did not change, compared to $NI \; (G=75)$ (Fig. \ref{fig:g220-ndna}a). We observe that: larger nDNA distances within species hinder speciation ($NI \; (G=220)$), while shorter distances between different species promotes speciation and hybridization ($\sigma_w=0.025$). Mitochondrial distances for $NI \; (G=220)$ were similar to $NI \; (G=75)$, supporting more diversity intraspecies.

\begin{figure}[h]
\begin{center}
        \includegraphics[width=0.6\linewidth]{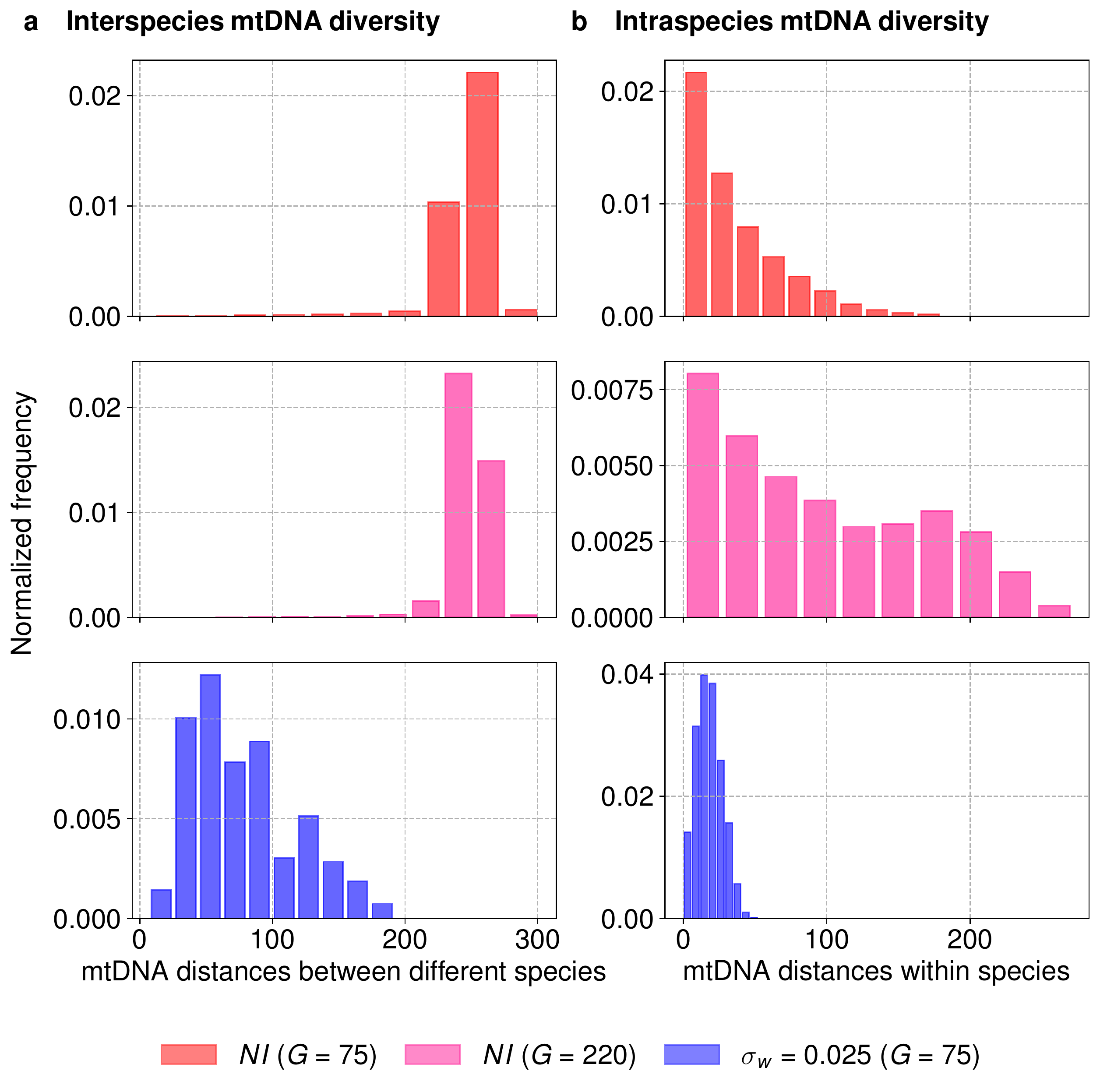}
\end{center}
	\caption{Mitochondrial distances between pair of individuals belonging to different species (a) and from the same species (b).}
	\label{fig:g220-mtdna}
\end{figure}

\pagebreak
\bibliographystyle{naturemag}
\bibliography{references}

\makeatletter\@input{file.tex}\makeatother